\documentclass[11pt]{article}
\usepackage[affil-it]{authblk}
\usepackage{color}
\usepackage{amsmath,amsfonts,amssymb,mathrsfs,amsthm}
\numberwithin{equation}{section}
\numberwithin{figure}{section}
\usepackage{graphicx}
\usepackage{listings}
\usepackage{enumerate,enumitem}
\usepackage{mathtools}
\usepackage[a4paper, total={6in, 8in}]{geometry}
\usepackage{cite}
\usepackage[colorlinks,linkcolor=blue,citecolor=blue]{hyperref}
\usepackage{empheq}
\usepackage{pdfpages}
\usepackage[titletoc]{appendix}
\usepackage[utf8]{inputenc}
\usepackage{xspace}
\usepackage{footmisc}

\newtheorem{defn}{Definition}[section]
\newtheorem{props}{Proposition}

\newcommand{\Figref}[1]{Fig.~\ref{#1}}
\newcommand{\Figsref}[1]{Figs.~\ref{#1}}
\newcommand{\R}{\mathcal{R}}
\newcommand{\C}{\mathcal{C}}

\newcommand{\Sectionref}[1]{Section~\ref{#1}}
\newcommand{\Sectionsref}[1]{Sections~\ref{#1}}
\newcommand{\kstar}{\overset{\star}{k}}
\newcommand{\pbar}{\bar{p}}

\newcommand{\Eqref}[1]{Eq.~\eqref{#1}}
\newcommand{\Defref}[1]{Def.~\ref{#1}}
\newcommand{\Eqsref}[1]{Eqs.~\eqref{#1}}
\newcommand{\Propref}[1]{Proposition~\ref{#1}}

\newcommand{\DecayO}[2]{O\left( \frac{1}{#1^{#2}} \right)}

\newcommand{\keyword}[1]{\emph{#1}}
\newcommand{\V}{\mathcal{V}}

\newcommand{\Sph}[1]{{\uppercase{#1}}}

\newcounter{mnotecount}[section]
\let\oldmarginpar\marginpar
\renewcommand\marginpar[1]{\-\oldmarginpar[\raggedleft\footnotesize #1]%
	{\raggedright\footnotesize #1}}

\title{Asymptotically hyperboloidal initial data sets from a parabolic-hyperbolic formulation of the Einstein vacuum constraints}
\author{F.~Beyer\footnote{Email: fbeyer@maths.otago.ac.nz},$\;$
	J. Ritchie\footnote{Email: jritchie@maths.otago.ac.nz}}
\affil{Department of Mathematics and Statistics, University of Otago, New Zealand.}
\begin{document}
	
	\maketitle
	
	\begin{abstract}
		In this paper we continue our investigations of R\'acz's parabolic-hyperbolic formulation of the Einstein vacuum constraints. Our previous studies of the asymptotically flat setting provided strong evidence for unstable asymptotics which we were  able to resolve by introducing a certain modification of R\'acz's parabolic-hyperbolic formulation. The primary focus of the present paper here is  the asymptotically hyperboloidal setting. We provide evidence through a mixture of numerical and analytical methods that the asymptotics of the solutions of R\'acz's parabolic-hyperbolic formulation are stable, and, in particular, no modifications are necessary  to obtain solutions which are asymptotically hyperboloidal. 
	\end{abstract}
	
	%\tableofcontents
	
	\section{Introduction}
	%% What are constraints? %%
	The triple $(\Sigma,\gamma_{ab},K_{ab})$ of a $3$-dimensional differentiable manifold $\Sigma$, Riemannian metric $\gamma_{a b}$ and smooth symmetric tensor field $K_{ab}$ on $\Sigma$ is called a \textit{vacuum initial data set} if it satisfies the \textit{vacuum constraint equations} 
	\begin{align}
		\prescript{(3)}{}{R}-K_{ab}K^{ab}+K^{2}=0,\quad \nabla_{a}{K^{a}}_{c}-\nabla_{c}K=0,
		\label{VacuumConstraints}
	\end{align} 
	everywhere on $\Sigma$, where $\nabla_{a}$ is the covariant derivative associated with $\gamma_{ab}$ and $\prescript{(3)}{}{R}$ is the corresponding Ricci scalar. Abstract tensor indices $a,b,\ldots$ are raised and lowered with the metric $\gamma_{ab}$, and $K={K^a}_a$. The constraint equations are a subset of the Einstein vacuum field equations (EFE). They earn the name \emph{constraints} as they place a restriction on the possible choices of initial data for the evolution equations obtained from EFE.

	%% Why do the constraints matter?
	Due to the pioneering work of Choquet-Bruhat and Geroch \cite{FouresBruhat:1952ji,ChoquetBruhat:1969cl} we know that if the constraint equations are satisfied on some initial surface then the evolution equations will ensure that they remain satisfied throughout the entire space-time. In fact, for every solution of the constraint equations there exists a unique maximal globally hyperbolic solution of EFE. Thus, in order to find solutions of the full Einstein vacuum equations, one must first seek  solutions of the constraint equations. However, solving the constraints can be difficult. A main reason is that the constraints are under-determined as they form a set of four equations for a total of twelve unknowns (counting each coordinate component of $\gamma_{ab}$ and $K_{ab}$, respectively). This means that some of the unknowns must be specified before the constraints can be solved. However, there is no geometrically or physically preferred way to decide which of the unknowns should be freely specifiable and which should be solved for.

	%% How does one solve the constraints?
	This property of the constraints (or really any under-determined system) can make the process of finding solutions challenging, as different choices of free data can lead to very different types of equations which in turn can produce solutions with very different properties. One of the most successful frameworks for solving the constraints is the conformal method introduced by Lichnerowicz and York (see \cite{bartnik2004, Baumgarte:2010vs} and references therein). In this approach, the constraints take the form of an elliptic system and are subsequently solved as a boundary value problem. 	
	The conformal method has been undeniable successful in the construction of solutions to the constraint equations \cite{bartnik2004}. It is not, however, without its limitations. Indeed, it is well known, for example, that the conformal method can fail if one seeks solutions of the constraints whose mean curvature is not close to constant (see \cite{dilts2017,anderson2018a} for an overview and references). Although there have been attempts to extend the conformal method, thereby removing this kind of issue, it is useful to explore other approaches to solving the constraints \cite{Bishop:1998cb,Matzner:1998hv,Moreno:2002dm,Bishop:2004gb}.

	%% Parabolic-hyperbolic approach
	A more recent alternative framework is to solve the constraint equations as a Cauchy problem \cite{Racz:2014kk,Racz:2014dx,Racz:2015gb,Racz:2015bu} instead of a boundary value problem. In his work \cite{Racz:2014kk}, R{\'a}cz suggests two evolutionary formulations of the constraint equations. The two formulations differ primarily in their treatment of the Hamiltonian constraint. We discuss this further in \Sectionref{Sec:ParabolicHyperbolicConstraints}. Solving the constraints as a Cauchy problem is interesting for a number of reasons. One of these is that, in general, one expects this approach to produce solutions with mean curvatures that are not necessarily close to constant. This reason alone makes this approach worth considering. However, it is not without its own pitfalls. The most obvious one is that Cauchy problems may in general not yield any control over the asymptotic behaviour of the solutions; this is clearly different for boundary value problems as a matter of principle. As a consequence physically relevant quantities, such as angular momentum or mass (see for example \cite{Szabados:2009ig,Cerebaum:2016}), may only be defined under very restrictive conditions. It is therefore possible that this method produces initial data sets that lack a clear physical meaning.

	%% Issues and previous work.
	We have previously studied R{\'a}cz's framework in the asymptotically flat setting in \cite{Beyer:2017tu,Beyer:2018HW, Beyer:2020HW}, both for R{\'a}cz's hyperbolic-algebraic formulation \cite{Beyer:2017tu} and for the parabolic-hyperbolic formulation \cite{Beyer:2018HW}. We found that solutions are generically not asymptotically flat. In \cite{Beyer:2020HW} we finally resolved this issue by proposing a small modification to R{\'a}cz’s original parabolic-hyperbolic formulation (as for R{\'a}cz’s original equations, these ``modified'' equations are equivalent to the Einstein vacuum constraints). These new equations preserve the parabolic-hyperbolic character of the PDEs, but at the same time also yield solutions with stable asymptotically flat asymptotics. In a similar spirit, a completely different modification was suggested in \cite{Csukas:2020} in an attempt to resolve the instabilities present in R{\'a}cz's algebraic-hyperbolic formulation.

	%% Aim of the current work.
	In this paper we now continue this line of research for the asymptotically hyperboidal setting. As in \cite{Beyer:2017tu,Beyer:2018HW, Beyer:2020HW}, we restrict our attention to foliations of $\Sigma$ where each $2$-surface is diffeomorphic to a $2$-sphere. This allows us to use the same numerical pseudo-spectral methods developed previously in \cite{Beyer:2017jw,Beyer:2014bu,Beyer:2016fc,Beyer:2009vw}. Interestingly we find that R{\'a}cz's original hyperbolic-parabolic formulation performs exceptionally well in this setting. In fact we provide evidence that no modifications are necessary here to obtain solutions with stable asymptotically hyperboloidal asymptotics. As in previous papers our main interest here is the \emph{asymptotic behaviour} at  infinity. In particular we do not study the strong-field regime properties of the resulting initial data sets in this paper at all. 
	
	%% Outline of the paper
	The paper is outlined as follows: In \Sectionref{Sec:ParabolicHyperbolicConstraints} we briefly summarise the framework of $2 + 1$-decompositions and introduce R\'{a}cz’s original parabolic-hyperbolic formulation of the vacuum Einstein constraints as well as Kerr-Schild-like data sets. \Sectionref{Sec:Backgrounds} is then devoted to the discussion of the asymptotics; we define the concept of asymptotic hyperbolicity and what it means for the $2 + 1$-quantities introduced in \Sectionref{Sec:ParabolicHyperbolicConstraints}. This section yields analytical evidence for our claims which we then support by numerics in \Sectionref{Sec:BinaryBlackHoles}.

	\section{Preliminary material}
	\label{Sec:ParabolicHyperbolicConstraints}
	\subsection{The $2+1$-decomposition and R\'acz's parabolic-hyperbolic formulation of the vacuum constraints}	
	\label{SubSec:Racz2Plus1Decomp}	
	We now discuss the framework of $2+1$-decompositions of initial data sets and R\'{a}cz's parabolic-hyperbolic formulation of the vacuum constraints. Further details can be found in \cite{Racz:2014kk,Racz:2014dx,Racz:2015gb,Racz:2015bu}. We use the same conventions as in \cite{Beyer:2018HW}.
	
	Consider an arbitrary initial data set $(\Sigma,\gamma_{ab},K_{ab})$, where as before, $\gamma_{a b}$ is a $3$-dimensional Riemannian metric and $K_{ab}$ is a smooth symmetric tensor field on $\Sigma$; at this stage this is not yet required to be a solution of the vacuum constraints.  Recall also that the Levi-Civita covariant derivative associated with $\gamma_{a b}$ is labelled $\nabla_{a}$.
	We suppose there exists a smooth function $\rho:\Sigma \rightarrow \mathbb{R}$ whose level sets $\mathcal{S}_\rho$ are smooth $2$-surfaces in $\Sigma$ such that the collection of all these surfaces is a foliation of $\Sigma$. This foliation yields a decomposition of the initial data set $(\Sigma,\gamma_{ab},K_{ab})$, in full analogy to the standard $3+1$-decomposition of spacetime as follows.
	If $t^a$ is a tangent vector in $\mathcal{S}_\rho$ then $t^{a}\nabla_{a}\rho=0$ and the unit co-normal of $\mathcal{S}_\rho$ is 
	\begin{align}
		N_{a}=A\nabla_{a}\rho,
		\label{eq:defNa}
	\end{align}
	where $A>0$ is the \textit{lapse}. 
	The first and second fundamental forms induced on each surface $\mathcal{S}_\rho$ are
	\begin{align}
		h_{ab}=\gamma_{ab}-N_{a}N_{b},
		\label{eq:defhdd}
	\end{align}
	and
	\begin{align}
		k_{ab}=-\frac{1}{2}\mathcal{L}_{N}h_{ab},
		\label{eq:defkdd}
	\end{align}
	respectively. 
	The covariant derivative associated with $h_{ab}$ is $D_{a}$.
	The tensor 
	\begin{align*}
		{h^ a}_{ b}={\delta^a}_{b}-N^{a}N_{b}
	\end{align*}
	is the map that projects any tensor field defined on $\Sigma$ orthogonally to a tensor field that is tangent to $S_\rho$. 	
	If the contraction of each index of a tensor field defined on $\Sigma$ with $N_{a}$ or $N^{a}$ is zero then we say that the field is \textit{intrinsic (to the foliation of surfaces $S_{\rho}$)}.  Contracting all
	indices of an arbitrary tensor field with ${h^a}_b$ yields  an intrinsic tensor field. In fact, any tensor can be uniquely decomposed into its intrinsic and orthogonal parts, in particular
	\begin{align}
		\label{eq:Kdec}
		K_{ab}=\kappa N_{a}N_{b}+N_{a}p_{b}+N_{b}p_{a}+q_{ab},
	\end{align}
	with
	\begin{align}
		\kappa=N^{a}N^{b}K_{ab},\quad p_{a}={h^c}_{a}N^{b}K_{cb},\quad q_{ab}={h^c}_{a}{h^d}_{b}K_{cd}.
		\label{Eq:Decompose_Kdd}
	\end{align}
	The field $q_{ab}$ is symmetric (i.e. $q_{ab}=q_{ba}$) and can be further decomposed into its trace $q$ and trace-free  $Q_{ab}$ parts (with respect to $h_{ab}$) as
	\begin{align}
		q_{ab}=Q_{ab}+\frac{1}{2}q h_{ab}, \;\; Q_{ab}h^{ab}=0,
		\label{Eq:Decompose_qdd}
	\end{align}          
	with the relations 
	\begin{align}
		q=h^{ab}q_{ab},\; Q_{ab}h^{ab}=0.
	\end{align}
	Note that $Q_{ab}$ is symmetric (i.e. $Q_{ab}=Q_{ab}$). 
	
	Now pick a vector field $\rho^a$ such that 
	\begin{equation}
		\rho^{a}\nabla_{a}\rho =1.
	\end{equation}    
	According to \Eqref{eq:defNa}  there must exist a unique intrinsic vector field $B^{a}$, called the \textit{shift}, such that
	\begin{equation}
		\rho^a = A N^{a} + B^{a}.
		\label{N_rho}
	\end{equation}
	Given \eqref{N_rho}, we can write \Eqref{eq:defkdd} as
	\begin{align}
		k_{ab}
		=-A^{-1}\left(\frac{1}{2}\mathcal{L}_{\rho}h_{ab}-D_{\left( a\right.}B_{\left. b\right)} \right)=:A^{-1}\kstar_{ab}.
		\label{eq:defkStar}
	\end{align}
	We also define
	\begin{align}
		\kstar:=h^{ab}\kstar_{ab}.
		\label{eq:defkStar2}
	\end{align}
	The Ricci scalar ${}^{(3)}R$ associated with $\gamma_{ab}$ can now be decomposed as
	\begin{align}
		{}^{(3)}R=&{}^{(2)}R-\left(A^{-2} \kstar^2 +A^{-2}\kstar_{ab}\kstar^{ab}+2A^{-1}D^{a}D_{a}A-2\left( A^{-1}\mathcal{L}_{N}\kstar-A^{-2}\mathcal{L}_{N}A \right)  \right),
		\label{Eq:R3}
	\end{align}
	where the Ricci scalar associated with the induced metric $h_{ab}$ is called ${}^{(2)}R$, and the \emph{intrinsic acceleration} is 
	\begin{align}
		v_{b}&=N^{a}\nabla_{a}N_{b}= -A^{-1}D_{b}A.
		\label{eq:defvb}
	\end{align}
	
	Finally, the \emph{Hawking mass} \cite{Alcubierre:Book} of a $\rho=\mathrm{const}$-surface
	\begin{align}
		\label{eq:hm1}
		m_{H}=\sqrt{\frac{| \mathcal{S}_\rho |}{16\pi}}\left( 1 + \frac{1}{16\pi}\oint_{\mathcal{S}_\rho}\Theta^{(+)}\Theta^{(-)}d\mathcal{S}  \right),
	\end{align}
	where $| \mathcal{S}_\rho |$ is the surface area of $\mathcal{S}_\rho$ and $\Theta^{(\pm)}$ are the in-and outgoing null expansion scalars defined with respect to suitably normalised future-pointing null normals of $\mathcal{S}_\rho$ when the initial data set is interpreted as an embedded hypersurface in a spacetime. These can be expressed as
	\begin{equation}
		\label{eq:hm2}
		\Theta^{(\pm)}=-\Bigl( q \pm A^{-1}\kstar \Bigr).
	\end{equation}

	Given all this, one can now decompose the vacuum momentum constraint \Eqref{VacuumConstraints} into their normal and intrinsic parts. According to \cite{Racz:2015gb} this, together with the Hamiltonian constraint, yields R\'acz's parabolic-hyperbolic formulation of the Einstein vacuum constraints: 
	\begin{align}
		\overset{\star}{k}\mathcal{L}_{\rho}A+A^2 D^{a}D_a A-\overset{\star}{k}B^a D_{a}A=\,
		&\frac 12 A^3 E
		+\frac 12A F,
		\label{ParabolicEquation_Racz}
		\\
		\mathcal{L}_{\rho}q-B^a D_{a}q-AD_{a}p^{a}-2p^{a}D_{a}A=\,&\overset{\star}{k} {}^{ab}Q_{ab}+\frac{1}{2}q\overset{\star}{k} -\overset{\star}{k}\kappa,
		\label{FinalSystemDiffNorm_Racz}
		\\
		\begin{split}
			\mathcal L_{\rho} p_c-B^a D_{a}p_{c}-\frac{1}{2}AD_{c}q-\kappa  D_c A +{Q^a}_{c}D_{a}A+\frac{1}{2}qD_{c}A=\,&p_a D_b B^a
			-AD_{a} {Q^a}_{c}
			\\
			&+\overset{\star}{k}p_c + AD_{c}\kappa,
		\end{split}
		\label{FinalSystemDiffMom_Racz}
	\end{align}
	\newcommand{\ParabolicHyperbolicR}{\Eqsref{ParabolicEquation_Racz}--\eqref{FinalSystemDiffMom_Racz}\xspace}
	where 
	\begin{align}
		\label{eq:defE}
		E&={}^{(2)}R+2\kappa q-2p^{a}p_{a}-Q_{ab}Q^{ab}+\frac{1}{2}q^2,
		\\
		\label{eq:defF}
		F&=2(\partial_{\rho}\overset{\star}{k}-B^a D_{a}\overset{\star}{k})-\overset{\star}{k}_{ab}{\overset{\star}{k}}^{ab}-\overset{\star}{k}^2.
	\end{align}	
	The structure of these equations suggest to group the various fields as follows:
	\begin{description}
		\item[Free data:] The fields $B_{a}$, $Q_{ab}$,  $h_{ab}$ and $\kappa$ are considered as {freely specifiable} in \ParabolicHyperbolicR{} everywhere
		on $\Sigma$. Notice from the above that $\kstar$, $D_a$,
		${}^{(2)}R$, $Q_{ab}$ and $F$ (and all other index versions of these
		intrinsic fields such as $Q^{a}{}_b$ etc.) are determined by the free data everywhere on
		$\Sigma$.
		\item[Unknowns:] The fields $A$, $q$ and $p_{a}$ are considered as the unknowns. Given free data, the task is to determine these as solutions of \ParabolicHyperbolicR{}. Observe here that all coefficients in these equations are determined by the free data everywhere on $\Sigma$.
		\item[Cauchy data:] Once free data have been specified, \ParabolicHyperbolicR{} are solved as a Cauchy problem  for the unknowns. The Cauchy data\footnote{\emph{Cauchy data} (or initial data) for $(q,A,p_{a})$ for the Cauchy problem of \ParabolicHyperbolicR{}  should not be confused with \emph{initial data sets} $(\gamma_{ab},K_{ab})$.}  for $A$, $q$ and $p_{a}$ are specified freely on an arbitrary $\rho=\rho_0$-surface of $\Sigma$. We always assume that the Cauchy data for $A$ are positive.
	\end{description}
	It was shown in  \cite{Racz:2014kk} that given smooth \emph{free data} everywhere $\Sigma$ with the property that the \keyword{parabolicity condition} 
	\begin{equation}
		\label{eq:parabolcond}
		\kstar<0
	\end{equation}
	holds everywhere on $\Sigma$,
	the \emph{Cauchy problem} of \ParabolicHyperbolicR{} in the \emph{increasing} $\rho$-direction is well-posed. This means that for arbitrary smooth \emph{Cauchy data} for $A$, $q$ and $p_a$ on an arbitrary $\rho=\rho_0$-leaf of the $2+1$-decomposition of $\Sigma$ the equations have a unique smooth solution $A$, $q$ and $p_{a}$ at least in a neighbourhood of the initial leaf.
	If the free data are such that $\kstar$ is positive instead, then the {Cauchy problem} in the \emph{decreasing} $\rho$-direction is well-posed.         
	In any case, \ParabolicHyperbolicR{} is therefore a quasilinear parabolic-hyperbolic system. It is important to notice that $\kstar$ is fully determined by the free data and \Eqref{eq:parabolcond} can therefore be verified \emph{prior} to solving \ParabolicHyperbolicR{}.

	For the rest of this paper we assume that the level sets of the function $\rho$ in our $2+1$-decomposition are diffeomorphic to the $2$-sphere $\mathbb S^2$. Because of this we assume that
	\[\Sigma=(\rho_{-},\infty)\times\mathbb S^2\]
	for some $\rho_{-}>0$, and we write the points in $\Sigma$ as $(\rho,p)$ with $\rho\in (\rho_{-},\infty)$ and $p\in\mathbb S^2$. Observe carefully that we often use the same symbol $\rho$ for the real parameter $\rho\in (\rho_{-},\infty)$ as well as for the \emph{function} $\rho$ defining the $2+1$-decomposition.
	Recall that all of the fields in \ParabolicHyperbolicR{} are smooth tensor fields on $\Sigma$, and, all of these fields are intrinsic to the $2$-sphere foliation (in the sense above). 
	Any intrinsic field on $\Sigma$ can be interpreted equivalently as
	a $1$-parameter family of fields on $\mathbb S^2$ defined by the
	$\rho$-dependent pull-back along the $\rho$-dependent map
	\begin{equation}
		\label{eq:Phirhomap}
		\Psi_\rho: \mathbb S^2\rightarrow\Sigma,\quad p\mapsto (\rho,p), 
	\end{equation}
	to $\mathbb S^2$. We use abstract indices  $A,B,\ldots$ to denote fields on $\mathbb S^2$ in order to distinguish them from fields on $\Sigma$ labelled with indices $a,b,\ldots$ as before. For example, the $\rho$-dependent pull-back of the intrinsic field $p_a$ on $\Sigma$ to $\mathbb S^2$ via $\Psi_\rho$ is labelled as $p_A$, or, when it is important to emphasise the dependence on $\rho$, as $p_A(\rho)$.
	It is easy to check that it is allowed to replace all indices $a,b,\ldots$ in \ParabolicHyperbolicR{} by $A,B,\ldots$  if, at the same time, each Lie-derivative along $\rho^a$ is replaced by the $\rho$-parameter derivative denoted as $\partial_\rho$. In most of this paper we shall indeed interpret \ParabolicHyperbolicR{} as evolution equations for $\rho$-dependent fields on $\mathbb S^2$ and therefore write the fields with indices $A,B,\ldots$.
	
	Following  \cite{Penrose:1984tf,Beyer:2015bv,Beyer:2014bu,Beyer:2016fc,Beyer:2017jw,Beyer:2017tu}, all ($\rho$-dependent or not) tensor fields on $\mathbb S^2$ can  be decomposed into quantities with well-defined \keyword{spin-weights} (see \Sectionref{Sec:SWSHstuff} in the appendix for a quick summary). We can also express the covariant derivative operator $D_\Sph a$ defined with respect to the intrinsic metric $h_\Sph{ab}$  as follows. Let $\hat D_\Sph a$ be defined with respect to the round unit-sphere metric $\Omega_\Sph {ab}$. Since the difference $D_\Sph a-\hat D_\Sph a$ can be expressed by some smooth intrinsic tensor field, and, according \Sectionref{Sec:SWSHstuff}, the covariant derivative operator $\hat D_\Sph a$ can be written in terms of the $\eth$- and $\eth'$-operators \cite{Penrose:1984tf}, the covariant derivative operator $D_\Sph a$ can be decomposed into $\eth$- and $\eth'$-components plus other smooth tensor fields on $\mathbb S^2$. Performing this for each term of  \ParabolicHyperbolicR{}, all terms end up with consistent well-defined spin-weights, and all terms are explicitly regular: Standard polar coordinate issues at the poles of the $2$-sphere are not present. The $\eth$- and $\eth'$-derivatives can be calculated by means of \Eqsref{eq:eths} and \eqref{eq:eths2} once all spin-weighted fields have been   expanded in terms of \keyword{spin-weighted spherical harmonics}.
	From the numerical point of view this gives rise to a (pseudo)-spectral scheme. Further details related to our implementation of this scheme can be found in \Sectionref{sec:numimpl} and \cite{Beyer:2017tu, Beyer:2018HW,Beyer:2020HW}.

	\subsection{Kerr-Schild-like data sets}
	\label{SubSec:Kerr_Shild_Like}

	In this subsection we introduce initial data sets that are of so-called \emph{Kerr-Schild-like form}. We do not yet require that these are solutions of the constraints. Initial data sets of this form were very useful in \cite{Beyer:2018HW, Beyer:2020HW} in the asymptotically flat setting. As with our previous works \cite{Beyer:2018HW, Beyer:2020HW} Kerr-Schild-like initial data sets form the basis of our numerical implementation in \Sectionref{Sec:BinaryBlackHoles}. In order to use them in the asymptotically hyperboloidal setting we generalise those now as follows.
	\begin{defn}
		\label{Def:KerrSchild}
		A data set $(\Sigma, \gamma_{ab}, K_{ab})$ is called \textit{Kerr-Schild-like} if $\Sigma=\mathbb{R}^3\backslash \overline B$ where $B$ is a  ball in $\mathbb{R}^3$  and there exists a smooth function $V:\Sigma\rightarrow \mathbb{R}$ with $V<1$, a smooth vector field $l_{a}$, and a symmetric tensor field $\dot{\gamma}_{ab}$ such that 
		\begin{align}
			\gamma_{ab}=\delta_{ab}-Vl_{a}l_{b},
			\quad
			K_{ab}=\frac{\sqrt{1-V}}{2}\left(  \nabla_{a}\left( Vl_{b} \right)+\nabla_{b}\left( Vl_{a} \right) -\dot{\gamma}_{ab} \right),
			\label{KerrSchild}
		\end{align}
		where $\delta_{ab}$ is the Euclidian metric on $\Sigma$, $\left( \delta^{-1} \right)^{ab}$ its inverse, and $l_a$ satisfies the condition 
		\begin{align}
			\left( \delta^{-1} \right)^{ab}l_{a}l_{b}=1.
		\end{align}
	\end{defn}
	
	In order to  discuss $2+1$-decompositions of Kerr-Schild-like initial data sets, we present some useful formulas. First, we define the vector $\tilde{l}^{a}$ as
	\begin{align}
		\tilde{l}^{a}=(\delta^{-1})^{ab}l_{b},
	\end{align}
	which yields the relationship
	\begin{align}
		\label{eq:lnorm}
		\tilde{l}^{a}l_a=(\delta^{-1})^{ab}l_a l_b=1.
	\end{align}
	The contravariant metric $\gamma^{ab}$ is then given as
	\begin{align}
		\gamma^{ab}=(\delta^{-1})^{ab}+\frac{V}{1-V}\tilde{l}^{a}\tilde{l}^{b},
	\end{align}
	and
	\begin{align}
		\label{eq:normalla}
		l^a=\gamma^{ab}l_b=\frac 1{1-V}\tilde{l}^{a},\quad l^al_a=\frac 1{1-V}.
	\end{align}
	
	\newcommand{\normall}{f}
	Suppose now we have chosen a  smooth function $\rho$ on $\Sigma$ with the properties discussed in \Sectionref{SubSec:Racz2Plus1Decomp} giving rise to a foliation $S$ in terms of level sets $S_\rho$ diffeomorphic to the $2$-sphere. We restrict to the case where $l_a$ is normal to $S_\rho$, i.e.,
	\begin{equation}
		\label{eq:specla}
		l_a=\pm\normall\nabla_a\rho,
	\end{equation}
	with
	\begin{equation}
		\label{eq:lanorm}
		\normall=\frac 1{\sqrt{(\delta^{-1})^{ab}\nabla_a\rho\nabla_b\rho}}.
	\end{equation}
	From \Eqsref{eq:defNa}, \Eqref{eq:specla} and \Eqref{eq:normalla} we find that
	\begin{equation}
		N_a=\sqrt{1-V}\, l_a,
	\end{equation}
	which means that the lapse defined in \Eqref{eq:defNa} is
	\begin{equation}
		\label{eq:KSA}
		A=\normall \sqrt{1-V}.
	\end{equation}
	It now follows from \Defref{Def:KerrSchild} and \Eqref{eq:defhdd} that
	\begin{equation}
		\label{eq:KShab}
		h_{ab}=\delta_{ab}-l_a l_b.
	\end{equation}
	
	Given adapted coordinates $(\rho,\vartheta,\varphi)$ on $\Sigma$, for which the vector $\rho^a$ in \Eqref{N_rho} has the representation $\partial_\rho^a$, the shift vector field $B^ a$ is determined by
	\begin{equation}
		\partial_\rho^a=\rho^a=A N^a+B^ a.
	\end{equation}
	It satisfies 
	\begin{equation}
		\label{eq:KSBa}
		B_ a=\rho^bh_{ab},
	\end{equation}
	and $k_{ab}$, $\overset{\star}{k}_{ab}$ and $\kstar$ can be calculated from \Eqsref{eq:defkStar} and \eqref{eq:defkStar2}.       
	Since
	\begin{equation}
		\label{eq:KSKab}
		K_{ab}
		=\frac{2-V}{4(1-V)}\left(\nabla_a V N_b+\nabla_b V N_a\right)
		+\frac{V}{2}\left(\nabla_a N_b+\nabla_b N_a\right)
		-\frac{\sqrt{1-V}}{2}\dot\gamma_{ab},
	\end{equation} 
	\Eqref{Eq:Decompose_Kdd} yields
	\begin{align}
		\label{eq:KSkappa}
		\kappa&=\frac{2-V}{2(1-V)^{3/2}} \tilde l^a\nabla_a V 
		-\frac{\sqrt{1-V}}{2}\dot\gamma_{ab}N^aN^b,
		\\
		\label{eq:KSpa}
		p_ a&=\frac{2-V}{4(1-V)} D_ a V 
		+\frac{V}{2}v_a -\frac{\sqrt{1-V}}{2}\dot\gamma_{cb}{h^{c}}_{a}N^b,
		\\
		\label{eq:KSqab}
		q_{ab}&=-V {k}_{ab}-\frac{\sqrt{1-V}}{2}\dot\gamma_{cd}{h^{c}}_{a}{h^{d}}_{b},
	\end{align}
	where $v_a$ is given by \Eqref{eq:defvb}. The quantities $q$ and $Q_ {ab}$ are then determined by
	\Eqref{Eq:Decompose_qdd}.

	\section{Asymptotically hyperboloidal data sets}
	\label{Sec:Backgrounds}
	
	Consider an arbitrary initial data set (not necessarily a solution of the vacuum constraints) $(\Sigma,\gamma_{ab},K_{ab})$, where $\Sigma = (\rho_-,\infty)\times\mathbb{S}^2$ for some $\rho_->0$, $\gamma_{ab}$ is a Riemann metric and $K_{ab}$ is a smooth symmetric $(0,2)$-tensor field as before. 
	It follows from \Sectionref{SubSec:Racz2Plus1Decomp} that this can equivalently be described by the collection of $1$-parameter fields $(A,\kappa,q,p_\Sph {a},B_\Sph {a},Q_\Sph {ab},h_\Sph{ab})$ on $\mathbb S^2$ given some function $\rho$. Because of this we shall often speak of $(A,\kappa,q,p_\Sph {a},B_\Sph {a},Q_\Sph {ab},h_\Sph{ab})$ as \emph{the $2+1$-fields associated with $(\gamma_{ab},K_{ab})$}, or vice versa, of $(\gamma_{ab},K_{ab})$ as the \emph{initial data set associated with the $2+1$ quantities} $(A,\kappa,q,p_\Sph {a},B_\Sph {a},Q_\Sph {ab},h_\Sph{ab})$. 
	
	\subsection{Asymptotic hyperbolicity}
	
	We start with a general definition of asymptotic hyperbolicity given in \cite{PhD:HypDef}. 
	\begin{defn}
		\label{def:hyperboloidal}
		Consider a smooth manifold $\Sigma$ with a Riemannian metric $\gamma_{ab}$ and smooth symmetric tensor field $K_{ab}$ (not necessarily a solution of the vacuum constraints). Then we call $(\Sigma,\gamma_{ab},K_{ab})$ asymptotically hyperboloidal if there exists a triple  $(\Lambda,\Omega,\psi)$ where
		\begin{enumerate}
			\item $\Lambda$ is a smooth manifold-with-boundary. 
			\item $\Omega:\Lambda \rightarrow \mathbb{R}$ is a smooth non-negative function which vanishes precisely on $\partial\Lambda$ but whose gradient $\boldsymbol{d}\Omega$ does not vanish on $\partial\Lambda$. 
			\item $\psi: \Lambda\setminus\partial\Lambda\rightarrow\Sigma$ is a diffeomorphism such that $\Omega^{2}\psi^{\star}(\gamma_{ab})$ is a Riemannian metric on $\Lambda\setminus\partial\Lambda$ which extends smoothly\footnote{The specific smoothness requirements depend on the application; as discussed below.\label{footnote:smoothness}} as a Riemannian metric to $\partial\Lambda$.
			\item The trace $K={K^{a}}_a$ of $K_{ab}$ with respect to $\gamma_{ab}$ is bounded away from zero near $\partial\Lambda$ when pulled back to $\Lambda$.
			\item Let $L_{ab}$ be the trace-free part of $K_{ab}$ and $L^{ab}=\gamma^{ac}\gamma^{bd}L_{cd}$. Then the field $\Omega^{-3}(\psi^{-1})_\star L^{ab}$ defined on $\Lambda\setminus\partial\Lambda$ extends smoothly\footref{footnote:smoothness} to $\partial\Lambda$. 
		\end{enumerate}
	\end{defn}

	\paragraph{Expansions and a minimal regularity characterisation of asymptotic hyperbolicity.}
	In order to analyse the asymptotics of initial data sets on $\Sigma=(\rho_-,\infty)\times\mathbb S^2$ at $\rho=\infty$ in the light of \Defref{def:hyperboloidal}, we first introduce some further terminology. Since the results presented in this paper here require us to be more precise than \cite{Beyer:2020HW}, this notation here differs slightly from the one given there.
	To this end, let $\Omega_\Sph {ab}$ be the round unit sphere metric on $\mathbb S^2$ with the coordinate representation $\mathrm{diag}(1,\sin^{2}\vartheta)$ in standard polar coordinates $(\vartheta,\varphi)$ on $\mathbb S^2$.
	Let $T(\rho)$ be an arbitrary $1$-parameter family of tensor fields on $\mathbb S^2$ of some given arbitrary rank\footnote{Since the tensor rank is arbitrary here we do not write any indices for this general discussion.} where the parameter $\rho$ is in $(\rho_-,\infty)$. For each fixed $\rho\in (\rho_{-},\infty)$, the tensor field $T(\rho)$ is therefore a section in some tensor bundle over $\mathbb S^2$. The set of smooth sections in this tensor bundle is referred to as $C^\infty(\mathbb S^2)$ (regardless of the rank of the tensor bundle under consideration). It is a standard fact that the metric $\Omega_\Sph {ab}$ on $\mathbb S^2$ induces a norm on this tensor bundle, with respect to which we define $C^0 ( (\rho_-,\infty), C^\infty(\mathbb S^2))$ as the set of all $1$-parameter families $T(\rho)$ of smooth sections over $\mathbb S^2$ which depend continuously on the parameter $\rho$ {pointwise} on $\mathbb S^2$. The compactness of $\mathbb S^2$ then implies continuity uniformly on $\mathbb S^2$ for each $\rho\in (\rho_-,\infty)$. Consequently, given an arbitrary integer $k\ge 0$ (or $k=\infty$), we define $C^k ( (\rho_-,\infty), C^\infty(\mathbb S^2))$ as the set  of all $1$-parameter families of smooth sections $T(\rho)$ over $\mathbb S^2$ which are $k$-times continuously differentiable with respect to $\rho$ pointwise on $\mathbb S^2$ (and therefore uniformly on $\mathbb S^2$) for each $\rho\in (\rho_-,\infty)$.
	
	Given an arbitrary $1$-parameter family $T(\rho)$ in $C^k ( (\rho_-,\infty), C^\infty(\mathbb S^2))$ as above, we say that the $1$-parameter family $\tilde T(t):=T(1/t)$ is a member of $C^k ( [0,1/\rho_-), C^\infty(\mathbb S^2))$ provided all of its $t$-derivatives up to order $k$  extend continuously to smooth sections over $\mathbb S^2$ at $t=0$ 
	pointwise on $\mathbb S^2$ (and therefore uniformly on $\mathbb S^2$). Notice here that $t=0$ corresponds to the limit $\rho\rightarrow\infty$.
	We also say that a $1$-parameter family $T(\rho)$ in $C^\infty ( (\rho_{-},\infty), C^\infty(\mathbb S^2))$ satisfies
	\[T(\rho)=O(\rho^{-\ell})\]
	in the limit $\rho\rightarrow\infty$ for some $\ell\in\mathbb R$ provided  $\hat T(t):=t^{-\ell}T(1/t)$ is a member of $C^0 ( [0,1/\rho_-), C^\infty(\mathbb S^2))$.
	If  $\ell$ is a non-negative integer, this is the case if and only if $\tilde T(t):=T(1/t)=t^\ell\hat T(t)$ is a member of $C^\ell ( [0,1/\rho_-), C^\infty(\mathbb S^2))$ according to Taylor's theorem.
	Finally, we say that a $1$-parameter family $T(\rho)$ in $C^k ( (\rho_-,\infty), C^\infty(\mathbb S^2))$ has an \emph{asymptotic radial expansion of order $\ell$ near $\rho=\infty$} for integers $\ell_0$ and $\ell$ with $\ell_0<\ell$ provided there are $T^{(\ell_0)},\ldots T^{(\ell-1)}\in C^\infty(\mathbb S^2)$ (of consistent tensor rank) -- the \emph{coefficients of the expansion} -- such that 
	\begin{equation}
		T(\rho)=\sum_{i=\ell_0}^{\ell-1} {T^{(i)}}\rho^{-i}+O(\rho^{-\ell}).
	\end{equation}

	Let us now use these concepts to express the conditions for asymptotically hyperboloidal initial data sets in \Defref{def:hyperboloidal} in terms of the asymptotics of the corresponding $2+1$-fields.

	\begin{props}[\textbf{Asymptotically hyperboloidal initial data sets}]
		\label{Result:Hyp_Decays}
		An initial data set $(\Sigma,\gamma_{ab}, K_{ab})$ with $\Sigma=(\rho_-,\infty)\times\mathbb S^2$ (not necessarily a solution of the vacuum constraints) is asymptotically hyperboloidal provided the corresponding $2+1$-fields satisfy the following properties
		\begin{align}
			\label{eq:result1First}
			A&=A^{(1)} \rho^{-1}+O(\rho^{-2}),\quad
			q=q^{(0)}+O(\rho^{-1}),\quad p_\Sph a=O(1),\\
			h_\Sph{ab}&=\rho^{2}\Omega_\Sph {ab}+O(\rho),\quad
			h^\Sph{ab}=\rho^{-2}(\Omega^{-1})^\Sph {ab}+O(\rho^{-3}),\\
			\label{eq:result1Last}
			B_\Sph a&=O(\rho^{-1}),\quad
			Q_\Sph {ab}=O(\rho),\quad
			2\kappa-q=O(\rho^{-1}),
		\end{align}
		for some strictly positive function $A^{(1)}\in C^\infty(\mathbb S^2)$ and some nowhere zero function $q^{(0)}\in C^\infty(\mathbb S^2)$.
		
	\end{props}
	We emphasise that the conditions in \Propref{Result:Hyp_Decays} are in general \emph{sufficient}, but not always \emph{necessary}, for asymptotic hyperbolicity as we discuss in the proof below.              
	Without further notice we shall assume in this paper that $\rho_{-}$ is always sufficiently large so that all the $2+1$-quantities given by the expansions in \Propref{Result:Hyp_Decays} have the required properties on the whole interval $(\rho_-,\infty)$; especially the lapse $A$ is then positive everywhere on $\Sigma$ since $A^{(1)}$ is positive.
	
	We recall that \Defref{def:hyperboloidal} does not specify which degree of regularity is required at infinity for asymptotic hyperbolicity. As we see in the proof, \Propref{Result:Hyp_Decays} guarantees only a \emph{minimal degree of regularity} at infinity. The proof of \Propref{Result:Hyp_Decays} however also makes it clear that higher degrees of regularity can be obtained by requiring that more derivatives of the fields extend to $\rho=\infty$ than the ones specified by \Eqsref{eq:result1First} -- \eqref{eq:result1Last}. Ultimatively, the issue of regularity for solutions of the Einstein vacuum equations is addressed by \Propref{res:smoothness}.
	We shall also see, for example in \Sectionref{sec:sphsymm}, that the minimal degree of regularity given by \Propref{Result:Hyp_Decays} can give rise to $\log\rho$-terms for generic solutions of the vacuum constraints in expansions around $\rho=\infty$. This can render the resulting initial data sets unphysical because physically meaningful quantities, like the Bondi mass, may not be defined. 
	
	%%- Result proof -%%
	\begin{proof}[Proof of \Propref{Result:Hyp_Decays}]
		It is convenient to introduce a coordinate system $(\rho,\vartheta,\varphi)$ on $\Sigma$ which is adapted to the $2+1$-foliation in the sense that the coordinate representation of the map $\Psi_\rho$ defined in \Eqref{eq:Phirhomap} is $(\vartheta,\varphi)\mapsto (\rho,\vartheta,\varphi)$.           
		Consider now the manifold-with-boundary $\Lambda = [0,1/\rho_{-})\times \mathbb{S}^2$ equipped with coordinates $(t,\vartheta,\varphi)$ with boundary $\partial \Lambda = \{ 0 \}\times\mathbb{S}^2\subset\Lambda$. The map $\psi:\Lambda \backslash \partial \Lambda \rightarrow \Sigma$ defined $(t,\vartheta,\varphi)\mapsto (1/t,\vartheta,\varphi)$ in terms of these coordinates is clearly a diffeomorphism. We define $\Omega = t$ and note that, as $t\rightarrow 0$, i.e., when we approach the boundary, we have $\Omega\rightarrow0$ while $\textbf{d}\Omega$ never vanishes on $\Lambda$. The first two conditions in \Defref{def:hyperboloidal} are therefore satisfied. Regarding the third condition, we find easily from \Eqsref{eq:defNa}, \eqref{eq:defhdd} and \eqref{N_rho} that the coordinate representation of $\Omega^{2}\psi^{\star}\gamma_{ab}$ is
		\begin{equation*}
			\begin{pmatrix}
				\frac{A^2+|B|^2}{t^2} & -B_\Sph \vartheta & -B_\Sph \varphi\\
				-B_\Sph \vartheta & t^2 h_\Sph{\vartheta\vartheta} & t^2 h_\Sph{\vartheta\varphi} \\
				-B_\Sph \varphi & t^2 h_\Sph{\vartheta\varphi} & t^2 h_\Sph{\varphi\varphi}
			\end{pmatrix},
		\end{equation*}
		where $|B|^2=h^{ab}B_a B_b$. The assumptions that $A=A^{(1)} t+O(t^2)$ for positive $A^{(1)}$, $h^\Sph{ab}=t^2(\Omega^{-1})^\Sph {ab}+O(t^3)$,  $h_\Sph{ab}=t^{-2}\Omega_\Sph {ab}+O(t^{-1})$ and $B_\Sph a=O(t)$ are therefore minimally sufficient (but not necessary) to satisfy condition $3$ in \Defref{def:hyperboloidal}.
		
		Turning our attention to conditions $4$ and $5$ in \Defref{def:hyperboloidal}, we first note that $K=\kappa + q$. This is bounded away from zero at $t=0$ since  $q^{(0)}+\kappa^{(0)}>0$ is part of the hypothesis. A slightly lengthy calculation reveals that it follows from \Eqsref{eq:defNa}, \eqref{eq:defhdd}, \eqref{eq:Kdec}, \eqref{Eq:Decompose_qdd} and \eqref{N_rho} that $\Omega^{-3}(\psi^{-1})_{\star}L^{ab}$ has the coordinate representation (the components marked with ``$\cdot$'' are obtained by symmetry)
		\begin{equation*}
			t^{-3}\begin{pmatrix}
				\frac 13t^3\frac{2\kappa-q}t \frac{t^2}{A^{2}}
				& \frac 13t^4\frac{2\kappa-q}t\frac{t^2}{A^2} \frac{B^\Sph \vartheta}{t^3}-t^3\frac{t}A \frac{p^\Sph \vartheta}{t^2}
				& \frac 13t^4\frac{2\kappa-q}t\frac{t^2}{A^2} \frac{B^\Sph \varphi}{t^3}-t^3\frac{t}A \frac{p^\Sph \varphi}{t^2}\\
				\cdot
				& \hat L^{11}
				& \hat L^{12}\\
				\cdot & \cdot & \hat L^{22}
			\end{pmatrix}
		\end{equation*}
		with
		\begin{align*}
			\hat L^{11}&=\frac 13t^5\frac{2\kappa-q}t \frac{t^2}{A^{2}} \frac{(B^\Sph \vartheta)^2}{t^6}
			-2t^4\frac{t}{A} \frac{B^\Sph \vartheta}{t^3} \frac{p^\Sph \vartheta}{t^2}
			+t^3\frac{Q^\Sph {\vartheta\vartheta}}{t^3}
			-\frac 16t^3\frac{2\kappa-q}t \frac{h^\Sph{\vartheta\vartheta}}{t^2},\\
			\hat L^{12}&=\frac 13t^5\frac{2\kappa-q}t \frac{t^2}{A^{2}} \frac{B^\Sph \vartheta B^\Sph \varphi}{t^6}
			-t^4\frac{t}{A} \frac{B^\Sph \vartheta}{t^3} \frac{p^\Sph \varphi}{t^2}
			-t^4\frac{t}{A} \frac{B^\Sph \varphi}{t^3} \frac{p^\Sph \vartheta}{t^2}
			+t^3\frac{Q^\Sph {\vartheta\varphi}}{t^3}
			-\frac 16t^3\frac{2\kappa-q}t \frac{h^\Sph{\vartheta\varphi}}{t^2},\\
			\hat L^{22}&=\frac 13t^5\frac{2\kappa-q}t \frac{t^2}{A^{2}} \frac{(B^\Sph \varphi)^2}{t^6}
			-2t^4\frac{t}{A} \frac{B^\Sph \varphi}{t^3} \frac{p^\Sph \varphi}{t^2}
			+t^3\frac{Q^\Sph {\varphi\varphi}}{t^3}
			-\frac 16t^3\frac{2\kappa-q}t \frac{h^\Sph{\varphi\varphi}}{t^2}.
		\end{align*}
		The hypothesis therefore guarantees that $\Omega^{-3}(\psi^{-1})_{\star}L^{ab}$ extends at least continuously to the boundary $t=0$. This is the minimal regularity  sufficient to satisfy condition~5 of \Defref{def:hyperboloidal}.          
		
	\end{proof}

	\paragraph{Consequences for Kerr-Schild-like data sets.}
	\label{SubSec:Consequences_for_Kerr-Schild-like_data_sets}
	Following on from \Sectionref{SubSec:Kerr_Shild_Like} we now list the consequences of \Propref{Result:Hyp_Decays} for Kerr-Schild-like data sets. To this end, we equip the manifold $\Sigma=(\rho_-,\infty)\times\mathbb S^2$  with the same canonical coordinates $(\rho,\vartheta,\varphi)$ as before, and in addition, with coordinates  $(r,\theta,\phi)$ related by some coordinate transformation of the form
	\begin{equation}
		\label{eq:KSr}
		r=\rho+\rho^{-1} R(\rho,\vartheta,\varphi),\quad \theta=\vartheta,\quad \phi=\varphi,
	\end{equation}
	for some so far arbitrary positive function $R$ with the property that $\tilde R(t,\vartheta,\varphi)=R(1/t,\vartheta,\varphi)$ extends to a map in $C^\infty([0,1/\rho_-),C^\infty(\mathbb{S}^2))$. This means that
	\begin{equation}
		\label{eq:KSr2}
		r=\rho+O(\rho^{-1}).
	\end{equation}
	Note that we intentially do not allow a $O(1)$-term in this expansion, see below. 
	The purpose of introducing these coordinates $(r,\theta,\phi)$ is to define the flat metric $\delta_{ab}$ in \Sectionref{SubSec:Kerr_Shild_Like} as 
	\begin{equation}
		\delta_{ab}=\nabla_a r \nabla_b r+r^2\nabla_a\theta\nabla_b\theta+r^2\sin^2\theta\nabla_a\phi\nabla_b\phi.
	\end{equation}
	The other freedoms to specify Kerr-Schild-like initial data sets are the scalar function $V$ and the symmetric $(0,2)$-tensor field $\dot\gamma_{ab}$ which we now  decompose as
	\begin{equation}
		\label{eq:dotgammadecomp}
		\dot\gamma_{ab}=\delta\kappa N_aN_b+2\delta p_ {(a} N_{b)}+\frac 12\delta q h_{ab}+\delta Q_ {ab}
	\end{equation}
	in analogy to \Eqsref{eq:Kdec}
	in terms of scalar fields $\delta\kappa$ and $\delta q$, a purely intrinsic field $\delta p_ a$ and a purely intrinsic trace free symmetric $(0,2)$-tensor field $\delta Q_ {ab}$. Assuming that the fields in \eqref{eq:dotgammadecomp} behave appropriately at $\rho=\infty$,
	it is
	straightforward to show using the formulas in \Sectionref{SubSec:Kerr_Shild_Like} that\footnote{Notice that $\sqrt{1-V}=O(\rho^{-1})$ in \Eqref{KerrSchild} if \Eqref{eq:VExp} holds.} the associated $2+1$-quantities have the expansions
	\begin{alignat*}{3}
		A&=\sqrt{\V}\rho^{-1}+O(\rho^{-2}),&\quad
		\kappa&=\frac 1{\sqrt{\V}}+O(\rho^{-1}),\\
		q&=\frac2{\sqrt{\V}}+O(\rho^{-1}),&
		p_\Sph a&=O(1),\\
		h_\Sph{ab}&=\rho^{-2}\Omega_\Sph {ab}+O(1),&
		B_\Sph a&=O(\rho^{-1}),\\
		Q_\Sph {ab}&=O(\rho),&&
	\end{alignat*}
	provided that
	\begin{equation}
		\label{eq:VExp}
		V(\rho)=1-\V\rho^{-2}+O(\rho^{-3})
	\end{equation}
	for an arbitrary $\V>0$, which for simplicity we assume to be a constant in this paper. \Propref{Result:Hyp_Decays} therefore implies that such a Kerr-Schild-like initial data set is asymptotically hyperboloidal. We remark that the main purpose of the condition \Eqref{eq:KSr2} is to guarantee that $h_{AB}-\rho^{-2}\Omega_{AB}$ is $O(1)$ as given above (as opposed to $O(\rho)$). This is especially useful for applications involving the stricter conditions of \Propref{res:smoothness} below.
	It also turns out to be useful to notice that
	\[\kappa=\frac 1{\sqrt{\V}}+O(\rho^{-2}),\quad Q_\Sph {ab}=O(1),\]
	if we assume that $O(\rho^{-3})$ is replaced by $O(\rho^{-4})$  in \Eqref{eq:VExp} (and if the fields in \Eqref{eq:dotgammadecomp}  decay appropriately fast at $\rho=\infty$).

	\subsection{Asymptotically hyperboloidal solutions of the vacuum constraints}
	\label{sec:sphsymm}

	\subsubsection{Spherically symmetric initial data sets}
	\label{SubSec:Hyperboloidal_Sphereical}
	%%
	%%- General discussion of spherically symmetric solutions - %%
	%%- solutions constructed from our approach. ----------------%%
	%%

	As a first step to analyse general solutions of \ParabolicHyperbolicR, we start off with the
	spherically symmetric case following the general strategy introduced in \cite{Beyer:2017tu,Beyer:2018HW, Beyer:2020HW}.
	First we pick a spherically symmetric background initial data set which satisfies the hypothesis of \Propref{Result:Hyp_Decays} and is therefore
	asymptotically hyperboloidal. We start with a
	Kerr-Schild-like data set defined by constants $M>0$ and $\lambda\in\mathbb R$, an arbitrary function $V(\rho)$ with $V(\rho)<1$ and $r=\rho$,
	\begin{equation}
		\label{GeneralFreeDataNew}
		h_{AB}=\rho^2\Omega_{AB},\quad
		B_A=0,\quad
		\kappa=-\frac{2 M (1-V)^2-\rho^2\partial_{\rho}V}{2 \rho (1-V)^{3/2}
			\eta(\rho\, ;M)}+\frac{\lambda}\rho,\quad
		Q_{AB}=0,
	\end{equation}
	and
	\begin{equation}
		\label{GeneralSolNew}
		A=\sqrt{1-V},\quad
		p_A=0,\quad
		q=\frac{2\, \eta(\rho\, ;M)}{\rho^2 \sqrt{1-V}},
	\end{equation}
	where 
	\begin{align}
		\eta(\rho\, ;M)= \sqrt{\rho ((\rho-2 M) V(\rho)+2 M)}
		\label{Eq:etaDef}
	\end{align}
	for $\rho>2M$. In the case $\lambda=0$ (which we mostly focus on), this initial data set is isometric to the data induced on some spherically symmetric slice in the Schwarzschild spacetime with mass $M>0$. It is therefore a solution of the vacuum constraints. According to \Sectionref{SubSec:Kerr_Shild_Like} we have
	\begin{align}
		\delta\kappa&=\frac{\rho\left(  (2-V)\eta(\rho\, ;M) -\rho \right)\partial_{\rho}V + 2 M (1-V)^2}{\rho (1-V)^2\eta(\rho\, ;M)}-\frac{2\lambda}{\sqrt{1-V}},
		\\
		\delta q&=-4\frac{\eta(\rho\, ;M) -\rho V}{\rho^2 {(1-V)}},
		\quad
		\delta Q_{AB}=0,\quad
		\delta p_A=0,
		%\end{split}
	\end{align}              
	see \Eqsref{eq:KSKab} and \eqref{eq:dotgammadecomp}. Since we want this data set to be well-defined for all large $\rho$, we impose the restriction $V(\rho)>0$ in addition to $V(\rho)<1$ above. The asymptotics of the function $V(\rho)$ at $\rho=\infty$ determine the character of this initial data set; in particular, if $V(\rho)$ satisfies \Eqref{eq:VExp}, then this background initial data set is asymptotically hyperboloidal.
	
	As mentioned above we focus on the spherically symmetric case in this subsection here.
	To this end we restrict to solutions of \ParabolicHyperbolicR{} with free data \eqref{GeneralFreeDataNew} where $p_\Sph a =0$ and $A$ and $q$ only depend on $\rho$. With this, \ParabolicHyperbolicR{} take the form
	\begin{align}
		\rho\partial_{\rho}q + q &= 2\kappa,
		\label{SpecialPDE_q}
		\\
		\rho\partial_{\rho}A-\frac{1}{2}A &= -\frac{\rho^2}{4}\left( \frac{2}{\rho^2} + 2\kappa q + \frac{1}{2}q^{2}  \right)A^{3},
		\label{SpecialPDE_A}
	\end{align}
	where $\kappa$ is given by \Eqref{GeneralFreeDataNew}.
	We remark that the parabolicity condition \Eqref{eq:parabolcond} is satisfied (but this fact is only relevant in the non-spherically symmetric PDE case and not for the ODE case here). In general, \Eqref{SpecialPDE_q} can be integrated as
	\begin{equation}
		\label{eq:qExplPre}
		q=\frac{\C}{\rho}+\frac{2}{\rho}\int \kappa(\rho)d\rho,
	\end{equation}
	where $\C$ is a free integration constant. Once $q$ has been found we can integrate \Eqref{SpecialPDE_A} as
	\begin{equation}
		\label{eq:AExplPre}
		A=\sqrt{\frac{\rho}{2(m-M) + \rho + \mathcal{F}(\rho)}},
		\quad
		\mathcal{F}(\rho) = \frac{1}{2}\int{\rho^2}\left(2\kappa + \frac{1}{2}q  \right)q\, d\rho
	\end{equation}
	where $m$ is another free integration constant.
	
	Such integrations cannot be performed explicitly unless we first specify the function $V(\rho)$. Anticipating the choices in \Sectionref{Sec:BinaryBlackHoles} for the asymptotically hyperboloidal case, we shall therefore now make the specific choice
	\[V(\rho)=1-\V\rho^{-2}\]
	for a so far arbitrary constant $\V>0$ in consistency with \Eqref{eq:VExp}. In this case, the function $\kappa$ takes the form
	\begin{equation}
		\kappa=\frac{\rho^3-\V  M}{\sqrt{\V } \rho^{3/2} \sqrt{2 \V  M+\rho^3-\V  \rho}}+\frac{\lambda}\rho,
	\end{equation}
	and we can perform the integration in \Eqref{eq:qExplPre} explicitly
	\begin{equation}
		\label{eq:solutionqinterm}
		q=\frac{\C \sqrt{\V } \sqrt{\rho}+{2 \sqrt{2 \V  M+\rho^3-\V  \rho}}}
		{\sqrt{\V } \rho^{3/2}}+2\lambda\frac{\ln \rho}{\rho}.
	\end{equation}
	This family of solutions agrees with the particular solution $q$ given in \Eqref{GeneralSolNew} in the special case $\C=0$ and $\lambda=0$. In order to explicitly perform the integration in \Eqref{eq:AExplPre}, we find it helpful to make a specific choice for $\V$ now. It turns out that if we choose
	\[\V=27 M^2,\]
	then the radicand appearing in both the formulas for $\kappa$ and for $q$ above factorises and, for $\lambda=0$, we get
	\begin{align}
		\label{eq:sphsolkappa}
		\kappa&=\frac{\rho^2 +3 M \rho+9 M^2}{ M\rho^{3/2} \sqrt{ 27(\rho+6 M)}}
		\\
		\label{eq:sphsolq}
		q&=\frac{\C}{\rho} + \frac{2(\rho-3 M)}{\sqrt{27} M\rho}\sqrt{1+\frac{6 M}{\rho}}
		\\
		\label{eq:sphsolA}
		A&=\sqrt{\frac{108M^{2}\sqrt{\rho+6M}\rho}{ \left( 4\rho^3 + 27 M^{2}\rho\C^2 + 216 M^{2}(M-m) \right)\sqrt{6M + \rho} + 12\sqrt{3}\left( \rho^{5/2} + 3 M\rho^{3/2} - 18 M^{2}  \right)\C }},
	\end{align}
	so long as $\rho>3M$. Although it is not immediately clear from the above expressions, one can show that these function $\kappa$, $q$ and $A$ extend smoothly to $\rho=\infty$.
	The function $A$ agrees with the one in \Eqref{GeneralSolNew} in the case $\C=0$ and $m=M$. Thanks to \Propref{Result:Hyp_Decays} we easily confirm that the resulting initial data sets are asymptotically hyperboloidal since
	\begin{align*}
		\kappa&=\frac{1}{\sqrt{27} M}+\frac{\sqrt{27} M}{2 \rho^2}++\DecayO{\rho}{3},
		\\
		A&=\frac{\sqrt{27} M}{\rho}-\frac{27 \C M^2}{2 \rho^2}+\DecayO{\rho}{3},
		\\
		q&=\frac{2}{\sqrt{27} M}+\frac{\C}{\rho}-\frac{\sqrt{27} M}{\rho^2}+\DecayO{\rho}{3}.
	\end{align*}
	
	We can also easily compute the Hawking mass $m_H$ from \Eqsref{eq:hm1} and \eqref{eq:hm2} and find
	\begin{equation}
		m_{H}=m.
	\end{equation}
	Given that these data sets are asymptotically hyperboloidal, the {Bondi mass} therefore agrees with the free parameter $m$.
	
	Finally let us comment on the role of the parameter $\lambda$. 
	In most of the discussion we are interested in $\lambda=0$. The point of allowing arbitrary values for $\lambda$ for some of the discussion above is to provide at least one explicit mechanism for  generating $\log\rho$-terms in the expansions of our solutions at $\rho=\infty$; see \Eqref{eq:solutionqinterm}. In order for our initial data sets to extend smoothly to infinity and physical quantities like the mass to be well-defined, such $\log\rho$-terms must not occur. Indeed, we shall find that the coefficient $\kappa^{(1)}$ in the expansion of $\kappa$ plays a general role and must in general be assumed to vanish in order to avoid $\log\rho$-terms (which corresponds to the case $\lambda=0$ in \Eqref{GeneralFreeDataNew} provided $V$ satisfies \Eqref{eq:VExp}). Notice carefully for example that $\kappa^{(1)}$ is not required to vanish in \Propref{Result:AsymAnal_Hyp_R} below (where we do not worry  about $\log\rho$-terms), but is required to vanish in \Propref{res:smoothness}, see \Eqref{eq:AHbackgroundkappa} (where the objective is to get rid of \emph{all} $\log\rho$-terms).

	\subsubsection{General solutions of R\'acz's parabolic-hyperbolic formulation of the vacuum constraints}
	\label{Sec:Solving_the_constraints_on_asymptotically_spherical_backgrounds}
	In this section we study the asymptotics of general solutions of the vacuum constraints without symmetry requirements  (especially not restricting to the specific choices in \Sectionref{SubSec:Hyperboloidal_Sphereical}) obtained by solving R\'acz's original parabolic-hyperbolic formulation in \Sectionref{Sec:ParabolicHyperbolicConstraints} for a large class of asymptotically hyperboloidal background data sets. 				
	The results in this section are purely formal in the sense that certain \emph{a-priori regularity  assumptions}  are made without rigorously proving the existence of solutions that satisfy these assumptions. A particular purpose of \Sectionref{Sec:BinaryBlackHoles} is to provide at least numerical justifications that these a-priori assumptions make sense.

	\paragraph{Minimal regularity.} We begin with a {minimal regularity} characterisation of asymptotically hyperboloidal \emph{vacuum} initial data sets in the light of \Propref{Result:Hyp_Decays}. 
	\begin{props}
		\label{Result:AsymAnal_Hyp_R}
		Pick a (sufficiently large) constant $\rho_->0$ and let $\Sigma$ be the manifold $(\rho_-,\infty)\times\mathbb S^2$. Consider a background initial data set (not necessarily a solution of the Einstein vacuum constraints) associated with arbitrary $2+1$ fields $(\kappa, B_\Sph {a},Q_\Sph {ab},h_\Sph{ab})$ on $\Sigma$ that satisfy the conditions of \Propref{Result:Hyp_Decays}, and, in addition, are such that $\kappa^{(0)}$ is either strictly positive or strictly negative.
                
		Let $(A,q,p_\Sph a)$ be an arbitrary smooth solution on $\Sigma$ of
		\ParabolicHyperbolicR{} given by the free data $(\kappa,B_\Sph {a},Q_\Sph {ab},h_\Sph{ab})$ satisfying the a-priori regularity assumptions that, (i), $A$ is strictly positive and has an asymptotic radial expansion of order $2$, and, (ii), $q$ and $p_\Sph a$ have asymptotic radial expansions of order $1$.
                
		Then
		\begin{equation}
			\label{eq:AsymAnal_Hyp_R}
			A^{(0)}=0,\quad A^{(1)}=\frac 1{|\kappa^{(0)}|},\quad
			q^{(0)}=2\kappa^{(0)},\quad p_\Sph a^{(0)}=0.
		\end{equation}    
		In particular, the resulting vacuum initial data set associated with the $2+1$ fields $(A,\kappa,q,p_\Sph {a},B_\Sph {a},Q_\Sph {ab},h_\Sph{ab})$ on $\Sigma$ is asymptotically hyperboloidal.                                        
		
	\end{props}

	Before we continue we remark that it is important to carefully distinguish the largely free choice of \emph{background} data set $(\kappa,B_\Sph {a},Q_\Sph {ab},h_\Sph{ab})$ from the resulting \emph{vacuum} initial data set $(A,\kappa,q, p_\Sph {a},B_\Sph {a},Q_\Sph {ab},h_\Sph{ab})$.
	The former determines the \emph{free data, but not necessarily the Cauchy data}, used to solve \ParabolicHyperbolicR{}. While the background data set is not necessarily a solution of Einstein's vacuum constraints, it is nevertheless asymptotically hyperboloidal as a consequence of \Propref{Result:Hyp_Decays}. The resulting data set $(A,\kappa,q, p_\Sph {a},B_\Sph {a},Q_\Sph {ab},h_\Sph{ab})$ is an actual solution of the vacuum constraints.  Making certain \emph{a-priori assumptions} about the solution $(A,q,p_\Sph a)$ of
	\ParabolicHyperbolicR{} as explained before we establish that \eqref{eq:AsymAnal_Hyp_R} follows, which, by \Propref{Result:Hyp_Decays}, then implies that the data set is asymptotically hyperboloidal. % As a consequence of minimal regularity, however, expansions of such data sets at $\rho=\infty$ may  contain log-terms and the Bondi mass may  not be defined. 
	% This issue of smoothness is addressed by \Propref{res:smoothness}.
	
	As a rough summary we conclude that \emph{R\'acz's parabolic-hyperbolic formulation \ParabolicHyperbolicR{} yields asymptotically hyperboloidal vacuum initial data sets from asymptotically hyperboloidal (in general non-vacuum) background initial data sets.} This is interesting because this is different in the asymptotically flat setting \cite{Beyer:2018HW} for Racz's formulation. In the asymptotically flat setting, it is \emph{the modified parabolic-hyperbolic formulation} proposed in \cite{Beyer:2020HW} that \emph{yields asymptotically flat vacuum initial data sets from asymptotically flat (in general non-vacuum) background initial data sets}.

	\begin{proof}[Proof of \Propref{Result:AsymAnal_Hyp_R}]
		Suppose that the hypothesis of \Propref{Result:AsymAnal_Hyp_R} holds. Using \Eqref{eq:defkStar}, we first find that $\kstar$ has asymptotic radial expansion
		\begin{align}
			\kstar = -2/\rho + O( \rho^{-2} ).
			\label{Eq:kstar_result}
		\end{align}
		Even though this is strictly speaking not relevant for this proof, we remark that
		the leading order is negative and  the parabolicity condition \Eqref{eq:parabolcond} is therefore satisfied for all sufficiently large $\rho$. 
		
		In order to prove the conclusions of \Propref{Result:AsymAnal_Hyp_R}, we now input the asymptotic radial expansions into \ParabolicHyperbolicR and sort all terms by powers of $\rho$. Each $\rho$-coefficient then yields an equation for the expansion coefficients of the unknowns.	The main observations relevant for this proof are as follows. \Eqref{FinalSystemDiffNorm_Racz} is satisfied in leading order (which turns out to be of order $\rho^{-1}$) if $q^{(0)}=2\kappa^{(0)}$. Similarly, \Eqref{ParabolicEquation_Racz} is satisfied at leading order (which turns out to be of order $1$) if $A^{(0)}=0$. \Eqref{FinalSystemDiffMom_Racz} holds at leading order (the $\rho^{-1}$-term) if $p_\Sph {a}^{(1)}=0$. At the next-to-leading order \Eqref{ParabolicEquation_Racz} is satisfied provided
		\begin{equation*}
			A^{(1)}\left( 1-\left( \kappa^{(0)}A^{(1)} \right)^2 \right)=0.
		\end{equation*}           
		Assuming that $A$ is positive
		we  choose $A^{(1)}=1/|\kappa^{(0)}|$. Given all this it
		now follows from \Propref{Result:Hyp_Decays} that
		the data set corresponding to $(A,\kappa,q, p_\Sph {a},B_\Sph {a},Q_\Sph {ab},h_\Sph{ab})$ is asymptotically hyperboloidal.
	\end{proof}

	\paragraph{The smooth case.}	
	It is convenient for the following discussion to  apply the parameter transformation $t=1/\rho$ with $t\in (0,T)$ where $T=1/\rho_-$ as above. This transforms the manifold  $\Sigma=(\rho_0,\infty)\times\mathbb S^2$ into the manifold $(0,T)\times\mathbb S^2$. The main concern of the following discussion is the limit $t\rightarrow 0$. Notice that as for Proposition~\ref{Result:AsymAnal_Hyp_R}, the following result draws conclusions from certain a-priori regularity assumptions for the solutions of the constraints near $t=0$ (i.e., $\rho=\infty$). Whether these assumptions hold for any solution is not known (however, we back our results up numerically in Section~\ref{Sec:BinaryBlackHoles}). The main thing we establish is that any solution that satisfies these a-priori regularity assumptions \emph{extends smoothly to $t=0$}. This is important because it means that such solutions are free of all $\log t$-terms in their expansions near $t=0$.

	\begin{props}
		\label{res:smoothness}
		Pick a (sufficiently small) constant $T>0$ and let $\Sigma$ be the manifold $\Sigma=(0,T)\times\mathbb S^2$.
		Consider a background initial data set (not necessarily a solution of the Einstein vacuum constraints) associated with $2+1$ fields $(\kappa,B_\Sph {a},Q_\Sph {ab},h_\Sph{ab})$ on $\Sigma$ satisfying
		\begin{align}
			\label{eq:AHbackgroundFirst}
			h_\Sph{ab}(t)&=t^{-2}\Omega_\Sph {ab}+\hat h_\Sph{ab}(t),\\
			h^\Sph{ab}(t)&=t^2(\Omega^{-1})^\Sph {ab}+t^4 \check h^\Sph{ab}(t),\\
			\label{eq:AHbackgroundkappa}
			\kappa(t)&=\kappa^{(0)}+t^2\hat\kappa(t),\\
			\label{eq:AHbackgroundLast}
			B_\Sph a(t)&=t\hat B_\Sph a(t),
		\end{align}
		where
		$\kappa^{(0)}>0$ is a constant, and, $\hat h_\Sph{ab}$, $\check h^\Sph{ab}$, $\hat\kappa$, $\hat B_\Sph a$ and $Q_\Sph {ab}$ are elements of $C^\infty([0,T),C^\infty(\mathbb S^2))$. Suppose also that
		$D_\Sph a Q_\Sph {bc} (\Omega^{-1})^\Sph {ab}=0$ at $t=0$.
		
		Let $(A,q,p_\Sph a)$ in $C^\infty((0,T),C^\infty(\mathbb S^2))$ be an arbitrary solution on $\Sigma$ of
		\ParabolicHyperbolicR{} (where the parameter $\rho$ is replaced by $t=1/\rho$) determined by free data $(\kappa,B_\Sph {a},Q_\Sph {ab},h_\Sph{ab})$ with the a-priori regularity assumptions that, (i), $q\in C^4([0,T), C^\infty(\mathbb S^2))$, (ii), $A$ is a strictly positive function in  $C^4([0,T), C^\infty(\mathbb S^2))$, and, (iii), $p_\Sph a\in C^3([0,T), C^\infty(\mathbb S^2))$.
                
		Then $q$, $A$ and $p_\Sph a$ are in $C^\infty([0,T),C^\infty(\mathbb S^2))$ and the vacuum initial data set associated with $(A,\kappa,q,p_\Sph {a},B_\Sph {a},Q_\Sph {ab},h_\Sph{ab})$ is asymptotically hyperboloidal with a finite Bondi mass
		\begin{equation}
			\label{eq:hatMBondiMass1}
			m=\lim_{t\rightarrow 0} m_H(t),
		\end{equation}
		where
		\begin{equation}
			\label{eq:hatMBondiMass2}
			m_H(t)=\frac{4+(\underline{q^2-A^{-2}\kstar^2})/t^2}{8t},
		\end{equation}
		using the notation  in \Eqref{eq:average}. 
	\end{props}
	
	% Before we continue we remark that it is important to carefully distinguish the largely free choice of \emph{background} data set $(\hat A,\kappa,\hat q,\hat p_\Sph {a},B_\Sph {a},Q_\Sph {ab},h_\Sph{ab})$ from the resulting \emph{vacuum} initial data set $(A,\kappa,q, p_\Sph {a},B_\Sph {a},Q_\Sph {ab},h_\Sph{ab})$.
	% The former determines the \emph{free data, but not necessarily the Cauchy data} used to solve \ParabolicHyperbolicR{}. The fields $\hat A$, $\hat q$ and $\hat p_\Sph {a}$ are in fact ignored. While the background data set is not necessarily a solution of Einstein's vacuum constraints, it is nevertheless asymptotically hyperboloidal as a consequence of \Propref{Result:Hyp_Decays}. The resulting data set $(A,\kappa,q, p_\Sph {a},B_\Sph {a},Q_\Sph {ab},h_\Sph{ab})$ is an actual solution of the vacuum constraints.  Making certain \emph{a-priori assumptions} about the solution $(A,q,p_\Sph a)$ of
	% \ParabolicHyperbolicR{} as explained before we establish that \eqref{eq:AsymAnal_Hyp_R} follows, which, by \Propref{Result:Hyp_Decays}, then implies that the data set is asymptotically hyperboloidal.
        
	As with \Propref{Result:AsymAnal_Hyp_R} we begin by remarking that it is important to distinguish the largely free choice of \emph{background} data set $(\kappa,B_\Sph {a},Q_\Sph {ab},h_\Sph{ab})$ from the resulting \emph{vacuum} initial data set $(A,\kappa,q, p_\Sph {a},B_\Sph {a},Q_\Sph {ab},h_\Sph{ab})$.
        The former determines the \emph{free data, but not necessarily the Cauchy data}, used to solve \ParabolicHyperbolicR{}.
	 While the background data set is not necessarily a solution of Einstein's vacuum constraints, it is again nevertheless asymptotically hyperboloidal as a consequence of \Propref{Result:Hyp_Decays}. Given the a-priori assumptions for the solutions, which are significantly stronger than for \Propref{Result:AsymAnal_Hyp_R},  it turns out that  the equations can be used in an iterative way to establish smoothness at $t=0$. The leading coefficients of the expansions of the solutions can be calculated explicitly in analogy to 
	\Eqref{eq:AsymAnal_Hyp_R}; see \Eqsref{eq:AHExpCOeffFirst} -- \eqref{eq:AHExpCOeffLast}. For all of this, the particular expansions of the fields in \eqref{eq:AHbackgroundFirst} -- \eqref{eq:AHbackgroundLast} as well as the strong a-priori assumptions for the unknowns are crucial to cancel all terms in \ParabolicHyperbolicR{} which would otherwise be too singular.
	Asymptotic hyperbolicity -- with infinite regularity in contrast to result for \Propref{Result:AsymAnal_Hyp_R} -- then follows again by
	\Propref{Result:Hyp_Decays}. In particular, $\log\rho$-terms must  not be present at $\rho=\infty$ and the Bondi mass must be well-defined. It fact this mass can be calculated by \Eqsref{eq:hatMBondiMass1} -- \eqref{eq:hatMBondiMass2}; but see also the additional discussion after the proof of \Propref{res:smoothness}. The Bondi mass could be expressed explicitly in terms of the expansion coefficients of the background fields and of the solution $(q,p_\Sph c,A)$, but the resulting formula  turns out to be too lengthy to write here. 
	
	It is interesting to notice that the condition $K=const$ in \cite{PhD:HypDef} for the construction of smooth vacuum asymptotically hyperboloidal data sets is unnecessary here.

	\begin{proof}[Proof of \Propref{res:smoothness}]
		Suppose the hypothesis of \Propref{res:smoothness} holds.
		According to Taylor's theorem,  we therefore have
		\begin{align}
			\label{eq:Tayloransatz1}
			q(t)&=q^{(0)}+ q^{(1)} t+q^{(2)} t^2 +q^{(3)} t^3+q^{(4)} t^4+w_0(t)t^4,
			\\
			p_\Sph c(t)&=p^{(0)}_\Sph{c}+p^{(1)}_\Sph{c} t+p^{(2)}_\Sph{c}t^2+p^{(3)}_\Sph{c}t^3+w_{1,\Sph{c}}(t) t^3,
			\\
			\label{eq:Tayloransatz3}
			A(t)&=A^{(0)}+A^{(1)}t+A^{(2)}t^2+A^{(3)} t^3+A^{(4)} t^4+w_{2}(t)t^4,
		\end{align}
		for some $W=(w_0,w_{1,c},w_2)$ in $C^\infty((0,T),C^\infty(\mathbb S^2))\cap C^0([0,T), C^\infty(\mathbb S^2))$ which vanishes in the limit $t\rightarrow 0$.
		
		Given this we now proceed as in \Propref{Result:AsymAnal_Hyp_R}: We input the  expansions \Eqsref{eq:Tayloransatz1}--\eqref{eq:Tayloransatz3} as well as \eqref{eq:AHbackgroundFirst} -- \eqref{eq:AHbackgroundLast} into \ParabolicHyperbolicR and sort all terms by powers of $t$. Each $t$-coefficient then yields an equation for the expansion coefficients of the unknowns. The resulting analysis is straightforward but lengthy and has been performed with computer algebra. We suppress the details of this calculation here but find that
		\begin{gather}
			\label{eq:AHExpCOeffFirst}
			p^{(0)}_\Sph{c}+p^{(1)}_\Sph{c} t + p^{(2)}_\Sph{c}t^{2}=\frac{\hat D_\Sph c q^{(1)}}{2\kappa^{(0)}} t + p^{(2)}_\Sph{c}t^{2},
			\\
			q^{(0)} + q^{(1)}t+q^{(2)}t^2=2\kappa^{(0)} + q^{(1)}t - 2\hat{\kappa}(0)t^{2},
			\\
			\label{eq:AHExpCOeffLast}
			\begin{split}
				A^{(0)}&+A^{(1)}t+A^{(2)}t^2+A^{(3)}
				t^3=\frac{1}{\kappa^{(0)}} t-\frac{q^{(1)}}{2(\kappa^{(0)})^2}t^2\\
				&+
				\frac{(q^{(1)})^2-2-{{2}}(\kappa^{(0)})^2{\hat D}_\Sph a \hat B_\Sph b(0) (\Omega^{-1})^\Sph {ab}+4\kappa^{(0)}\hat\kappa(0){{-2(\kappa^{(0)})^2\hat h_\Sph{ab}(0)(\Omega^{-1})^\Sph {ab}}} }{4(\kappa^{(0)})^3}t^3.
			\end{split}
		\end{gather}
		Here, ${\hat D}$ is the covariant derivative defined with respect to the (by definition $t$-independent) metric $\Omega_\Sph {ab}$ on $\mathbb S^2$. We shall not write down the lengthy expressions for $q^{(3)},q^{(4)}$ and $p^{(3)}_\Sph{c}$ here for brevity. Notice that
		the quantities
		\[q^{(1)}, p^\Sph {(2)}_{c}, A^{(4)}\in C^\infty(\mathbb S^2)\]
		turn out to
		represent the asymptotic degrees of freedom of the space of all solutions -- the \emph{asymptotic data}. In particular, all the expansion coefficients of solutions can be written in terms of these asymptotic data in conjunction with the free data.
		
		It follows immediately from \Propref{Result:Hyp_Decays} that the corresponding initial data set is asymptotically hyperboloidal (with at least minimal regularity). In order to establish the claimed smoothness property, we next need to construct $t$-derivatives of arbitrary order at $t=0$ from the equations.
		Without going into the details of the lengthy calculations, it turns out that the equations
		can be written in the following schematic form
		\begin{equation}
			\label{eq:AHRemEq}
			\begin{split}
				\partial_t W(t,p)=&\frac 1t\text{diag}(-3,-1,0) W(t,p)\\
				&+H(t,p,q^{(1)}(p), p^{(2)}_\Sph{c}(p),A^{(4)}(p),W(t,p),{\hat D} W(t,p),{\hat D}^2 W(t,p))
			\end{split}
		\end{equation}
		for every $t\in (0,T)$ and $p\in \mathbb S^2$. Here, $H$ is a (lengthy, but explicitly known) function which is smooth in each of its arguments, especially at $t=0$.
		The fact that we can write the equations in this schematic form is the precise reason which allows us to draw our conclusions about smoothness as we demonstrate below.
		Without the specific a-priori regularity assumptions for $q$, $p_\Sph c$ and $A$ summarised in \Eqsref{eq:Tayloransatz1} -- \eqref{eq:Tayloransatz3} and without the specific assumptions on the free background fields and the related fields defined by \Eqsref{eq:AHbackgroundFirst} -- \eqref{eq:AHbackgroundLast}, \Eqref{eq:AHRemEq} would contain disastrous additional singular terms at $t=0$. These additional terms   would in general generate $\log t$-terms of arbitrary order at $t=0$. \emph{With} all these assumptions, however, these terms all cancel precisely. In fact, they cancel precisely \emph{independently of the particular values of the asymptotic data $q^{(1)}$, $p^{(2)}_\Sph{c}$ and $A^{(4)}$}. This strongly suggests that  smoothness is indeed a  property  of generic solutions.
		
		Now, since by assumption the field $W$ is in $C^0([0,T),C^\infty(\mathbb S^2))$ and, especially, vanishes at $t=0$, we can write \Eqref{eq:AHRemEq} in integral form
		\begin{equation}
			\label{eq:AHRemSol}
			\begin{split}
				W&(t,p)=\text{diag}(t^{-3},t^{-1},1)\\
				&\times\int_0^t \text{diag}(s^3,s,1)H(s,p,q^{(1)}(p), p^{(2)}_\Sph{c}(p),A^{(4)}(p),W(s,p),{\hat D} W(s,p),{\hat D}^2 W(s,p))ds.
			\end{split}
                      \end{equation}
                      Observe here that the integrand is continuous over the whole integration domain, including, most importantly $s=0$. 
		Given this, we proceed with the following inductive argument. Let us make the inductive assumption that we have shown that the solution $W$ of \Eqref{eq:AHRemSol} exists and that $W\in C^k([0,T),C^\infty(\mathbb S^2))$ for some arbitrary $k\ge 0$; the base  case $k=0$ for this inductive argument is a direct consequence of this hypothesis.  Given the regularity of the integrand we are allowed to use the substitution $s=t\tau$ which leads to
		\begin{equation*}
			\begin{split}
				&\frac{W(t,p)}t=\\
				&\times\int_0^1 \text{diag}(\tau^3,\tau,1)H(t\tau,p,q^{(1)}(p), p^{(2)}_\Sph{c}(p),A^{(4)}(p),W(t\tau,p),{\hat D} W(t\tau,p),{\hat D}^2 W(t\tau,p))d\tau.
			\end{split}
		\end{equation*}
		We conclude from this that not only $W$ itself but also $t^{-1} W(t)$ can be extended to an element of  $C^k([0,T),C^\infty(\mathbb S^2))$. \Eqref{eq:AHRemEq} then implies that the same is true for the $1$-parameter family of fields $\partial_t W$. We have therefore established that $W$ extends to an element of $C^{k+1}([0,T),C^\infty(\mathbb S^2))$. Since $k$ was arbitrary, we have therefore  established that $W$ is an element of $C^{\infty}([0,T),C^\infty(\mathbb S^2))$ as required.
		We remark that in all of these steps we have repeatedly used the fact that $\mathbb S^2$ is compact implicitly. 
		
		Let us now address the final claim regarding the Bondi mass. Since the resulting initial data set is asymptotically hyperboloidal with $C^\infty$-regularity at infinity (represented by $t=0$), the Bondi mass $m$ is the limit of the Hawking mass $m_H(t)$ at $t=0$ \cite{Penrose:1984tf,penrose1986}.
		Given \Eqref{eq:AHbackgroundFirst}, \Eqsref{eq:hm1} and \eqref{eq:hm2} yield
		\begin{align*}
			m_{H}(t)
			&=\frac{1+O(t)}{32\pi} \oint_{\mathcal{S}_\rho}\frac{4+\frac{q^2-A^{-2}\kstar^2}{t^2}}{t}\, t^2{d\mathcal{S}}+O(t),
		\end{align*}
		where ${d\mathcal{S}}$ represents the (in general $t$-dependent) volume form associated with the (in general $t$-dependent) metric $h_{AB}$. Using \Eqref{eq:AHbackgroundFirst} again, we conclude that the Hawking mass $m_H(t)$  has a finite limit at $t=0$ (which agrees with the Bondi mass $m$) provided
		\begin{equation}
			\label{eq:bondimassexistence}
			\underline{q^2-A^{-2}\kstar^2}=-{4}t^2+O(t^3), 
		\end{equation}
		using \Eqref{eq:average}, and
		\[m_{H}=m+O(t).\]
		In order to show that \Eqref{eq:bondimassexistence} holds for the class of initial data sets here, we observe that
		\Eqsref{eq:defkStar}, \eqref{eq:defkStar2} and
		\eqref{eq:AHbackgroundFirst} -- \eqref{eq:AHbackgroundLast}, together with the hypothesis that $\hat h_\Sph{ab}$, $\check h^\Sph{ab}$, $\hat\kappa$, $\hat B_\Sph a$ and $Q_\Sph {ab}$ are in $C^\infty([0,T),C^\infty(\mathbb S^2))$,  yield
		\begin{equation}
			\label{eq:kstarexp}
			\kstar=-2t+\Bigl((\Omega^{-1})^\Sph {ab}\hat D_\Sph a\hat{B}_\Sph{b}(0)
			+2\hat h_\Sph{ab}(0)(\Omega^{-1})^\Sph {ab}\Bigr)t^3+O(t^4).
		\end{equation}
		Using this together with the expressions for $q^{(0)}$,  $q^{(1)}$, $q^{(2)}$, $A^{(0)}$,
		$A^{(1)}$, $A^{(2)}$ and $A^{(3)}$ from \Eqsref{eq:AHExpCOeffFirst} --
		\eqref{eq:AHExpCOeffLast} yields 
		\begin{align*}
			q^2(t)-A^{-2}(t)\kstar{}^2(t)
			=-4t^2+O(t^3),
		\end{align*}		
		as required by \Eqref{eq:bondimassexistence}. The (explicitly known but lengthy) coefficient of the $O(t^3)$-term here
		gives an explicit formula for the Bondi mass $m$.
	\end{proof}
	
	\paragraph{Evolution of the Hawking mass and the Bondi mass.}
	While the Bondi mass can in principle be calculated by
	\Eqsref{eq:hatMBondiMass1} -- \eqref{eq:hatMBondiMass2} under the hypothesis of \Propref{res:smoothness} once the constraint equations are
	solved, we find in our numerical studies that numerical errors render practical
	calculations of the limit in \eqref{eq:hatMBondiMass1} impossible. As we demonstrate in
	\Sectionref{Sec:BinaryBlackHoles}, the following alternative approach provides a remedy for this. To this end we first notice that we can write \Eqref{ParabolicEquation_Racz} as
	\begin{equation*}
		-t\partial_t A-\frac 12 A+\frac{\rho^2}{4}\left( \frac{2}{\rho^2} +
		2\kappa q + \frac{1}{2}q^{2}  \right)A^{3}=\frac 12 A F_{[A]}
	\end{equation*}
	with
	\begin{equation}
		\label{eq:FA}
		F_{[A]}=
		-\frac{2}{t\overset{\star}{k}}A D^\Sph {a}D_\Sph a A
		+\frac{2}{A t} B^\Sph a D_\Sph {a}A
		+\left(\frac{1}{t\overset{\star}{k}} E+ 1 + \frac
		{\kappa q}{t^2} + \frac{1}{4t^2}q^{2}  \right) A^2
		+\left(\frac{1}{t\overset{\star}{k}} F-1\right) ,
	\end{equation}
	where we notice from \Eqref{SpecialPDE_A} that $F_{[A]}=0$ in the
	spherically symmetric case  where, in particular, $\kstar=-2t$.
	Secondly we find from \Eqref{FinalSystemDiffNorm_Racz} similarly that
	\begin{equation*}
		-t \partial_{t}q+q-2\kappa=\frac 12 F_{[q]}
	\end{equation*}
	with 
	\begin{equation}
		\label{eq:Fq}
		F_{[q]}=\frac 2{t} B^\Sph a D_\Sph {a}q+\frac 2{t} AD_\Sph {a}p^\Sph {a}+\frac 4{t} p^\Sph {a}D_\Sph {a}A
		+\frac 2{t}\,\overset{\star}{k} {}^{ab}Q_\Sph {ab}
		+2\left(\frac 1{2t}\overset{\star}{k}+1\right)(q-2\kappa),
	\end{equation}
	where \Eqref{SpecialPDE_q} implies that $F_{[q]}=0$ in the spherically symmetric case. According to \Eqref{eq:hatMBondiMass2}, the main quantity to determine the Bondi mass is
	\begin{equation}
		\label{eq:defaa}
		a:=q^2-\kstar^2A^{-2},
	\end{equation}
	for which a straightforward calculation yields 
	\begin{equation*}
		-t \partial_{t}a+3a+\kstar^2
		=F_{[a]}
	\end{equation*}
	with
	\begin{equation}
		\label{eq:defFa}
		F_{[a]}=\kstar^2 F_{[A]} A^{-2}
		+F_{[q]} q
		+2(t\partial_{t}\kstar-\kstar) \kstar A^{-2}  
		-\frac{\kstar^2-4t^2}{4t^2}(4\kappa+q)q.
	\end{equation}
	As before we notice that $F_{[a]}=0$ in the spherically symmetric case where $\kstar=-2t$ and $F_{[A]}=F_{[q]}=0$. According to \Eqsref{eq:hatMBondiMass2} and \eqref{eq:defaa},
	\begin{equation}
		\label{eq:MevolDef}
		m_H(t)=\frac 1{2t}+\frac{\underline a(t)}{8t^3}.
	\end{equation}
	This satisfies the differential equation
	\begin{equation}
		\label{eq:Mevol}
		\partial_t m_H=\frac{\underline{\kstar^2}-4t^2-\underline{F_{[a]}}}{8t^4},
	\end{equation}
	which can be readily solved as
	\begin{equation}
		\label{eq:Mevolsol}
		m_H(t)=m_H(T_0)+\int_{T_0}^t\frac{\underline{\kstar^2}(s)-4s^2-\underline{F_{[a]}}(s)}{8s^4} ds.
	\end{equation}
	
	Given now a sufficiently smooth initial data set $(A,\kappa,q,p_\Sph {a},B_\Sph {a},Q_\Sph {ab},h_\Sph{ab})$ with the property that
	\begin{equation}
		\label{eq:massconvergencecond}
		\underline{F_{[a]}}(t)=O(t^4)
	\end{equation}
	at $t=0$,
	it follows that $m_H(t)$ has a finite limit, i.e., the Bondi mass $m$,
	\begin{equation*}
		m=m_H(T_0)+\int_{T_0}^0\frac{\underline{\kstar^2}(s)-4s^2-\underline{F_{[a]}}(s)}{8s^4} ds
	\end{equation*}
	according to \Eqref{eq:hatMBondiMass1}.
	In saying this, we assume that the fields $\hat h_\Sph{ab}$, $\check h^\Sph{ab}$, $\hat\kappa$, $\hat B_\Sph a$ and $Q_\Sph {ab}$ defined by \Eqsref{eq:AHbackgroundFirst} -- \eqref{eq:AHbackgroundLast}  are in $C^\infty([0,T),C^\infty(\mathbb S^2))$ here and in all of what follows so that $\kstar^2-4t^2=O(t^4)$ as a consequence of \Eqref{eq:kstarexp}. The calculations in the proof of \Propref{res:smoothness} can be used to show that \Eqref{eq:massconvergencecond} holds under the conditions of \Propref{res:smoothness}. Interestingly, it turns out that \Eqref{eq:massconvergencecond} holds even under weaker conditions when the initial data set is therefore not necessarily fully smooth at infinity. In particular we can check by straightforward (but lengthy) calculations that this is the case provided the fields $\hat h_\Sph{ab}$, $\check h^\Sph{ab}$, $\hat\kappa$, $\hat B_\Sph a$ and $Q_\Sph {ab}$ defined by \Eqsref{eq:AHbackgroundFirst} -- \eqref{eq:AHbackgroundLast}  are in $C^\infty([0,T),C^\infty(\mathbb S^2))$ and provided: (i) $(q,p_\Sph c,A)$ are in $C^\infty((0,T),C^\infty(\mathbb S^2))$, and (ii), $q\in C^2([0,T), C^\infty(\mathbb S^2))$, $A$ is a strictly positive function in  $C^3([0,T), C^\infty(\mathbb S^2))$ and $p_\Sph c\in C^2([0,T), C^\infty(\mathbb S^2))$, and (iii), $p^\Sph {(0)}_c$, $p^\Sph {(1)}_c$, $A^{(0)}$ , $A^{(1)}$ , $A^{(2)}$, $q^{(0)}$ and $q^{(1)}$ have the values given in \Eqsref{eq:AHExpCOeffFirst} -- \eqref{eq:AHExpCOeffLast}.
	
	For practical calculations the idea is therefore to approximate $m$ by evolving $m_H(t)$ by means of \Eqref{eq:Mevol}. Since this means that we need to determine $F_{[a]}$ at every time step of the evolution, it  makes sense to evolve the combined system \ParabolicHyperbolicR{} and \eqref{eq:Mevol} for the combined set of unknowns ($q, p_A, A, m_H)$ simultaneously.
	Notice that $m_H(T_0)$ in \Eqref{eq:Mevolsol} is determined from the initial data of the quantity $m_H$ determined from \Eqsref{eq:defaa} and \eqref{eq:MevolDef} at $t=T_0$.

	\section{Numerical investigations}
	\label{Sec:BinaryBlackHoles}
	\subsection{Binary black hole background data sets}
	\label{Sec:A_superposition_method_for_generating_binary_black_hole_data}
	
	Before we present numerical solutions of the constraint equations and thereby attempt to confirm the theoretical results of the previous sections,
	we first present a summary of a binary black hole background initial data model. This was introduced  in \cite{Beyer:2018HW} where the reader can find a more in-depth discussion. The main idea is to use a binary black hole initial data set obtained by a Kerr-Schild superposition procedure as a background initial data set. As before this then determines the \textit{free data} (and in some circumstances also the \textit{Cauchy data}) to solve
	\ParabolicHyperbolicR{} as a Cauchy problem. Since such a background data set represents a binary black hole system, the hope is that the corresponding vacuum initial data set obtained in this way also represents initial data for a binary black hole system. 
	
	As in \cite{Beyer:2018HW,Beyer:2020HW}, the background initial data set is constructed using the formalism in \Sectionref{SubSec:Kerr_Shild_Like}. Owing to the symmetry of by assumption non-spinning binary black hole systems we restrict to the case in which $l_{a}$ is orthogonal to the level sets $S_\rho$ of the function $\rho$ in consistency with \Eqref{eq:specla}. Given two black hole mass parameters $M_+>0$ and $M_-\ge 0$ and two position parameters $Z_+$ and $Z_-$, we define this function $\rho$ as
	\begin{equation}
		\rho=(M_++M_-)\left(\frac{M_+}{\sqrt{x^2 +y^2 +(z-Z_+)^2}}+\frac{M_-}{\sqrt{x^2 +y^2 +(z+Z_-)^2}}\right)^{-1}.
		\label{Eq:rho_def}
	\end{equation}
	A particular property of this function is that $\rho\rightarrow r$ at $r\rightarrow\infty$ where
	\begin{equation}
		r = \sqrt{x^2 + y^2 + z^2}.
		\label{Eq:r_def}
	\end{equation}
	As in \cite{Beyer:2020HW} (but in contrast to \cite{Beyer:2018HW}), we also impose the \emph{centre of mass condition}
	\begin{equation}
		Z_+ M_{+}-Z_{-}M_{-}=0,
	\end{equation}
	which allows us to write
	\begin{equation}
		Z_{-}=Z,\quad Z_+ = \frac{M_-}{M_+}Z
		\label{Eq:Z_choices}
	\end{equation}
	for a single separation parameter $Z\ge 0$.
	An important consequence of this centre of mass condition is that $\rho$ satisfies
	\begin{equation}
		\rho=r+\DecayO{r}{},
		\label{Eq:rhoDecay}
	\end{equation}
	while without this condition we would have $\rho=r+O(1)$ in general.

\begin{figure}
		\centering
		\includegraphics[width=0.8\linewidth]{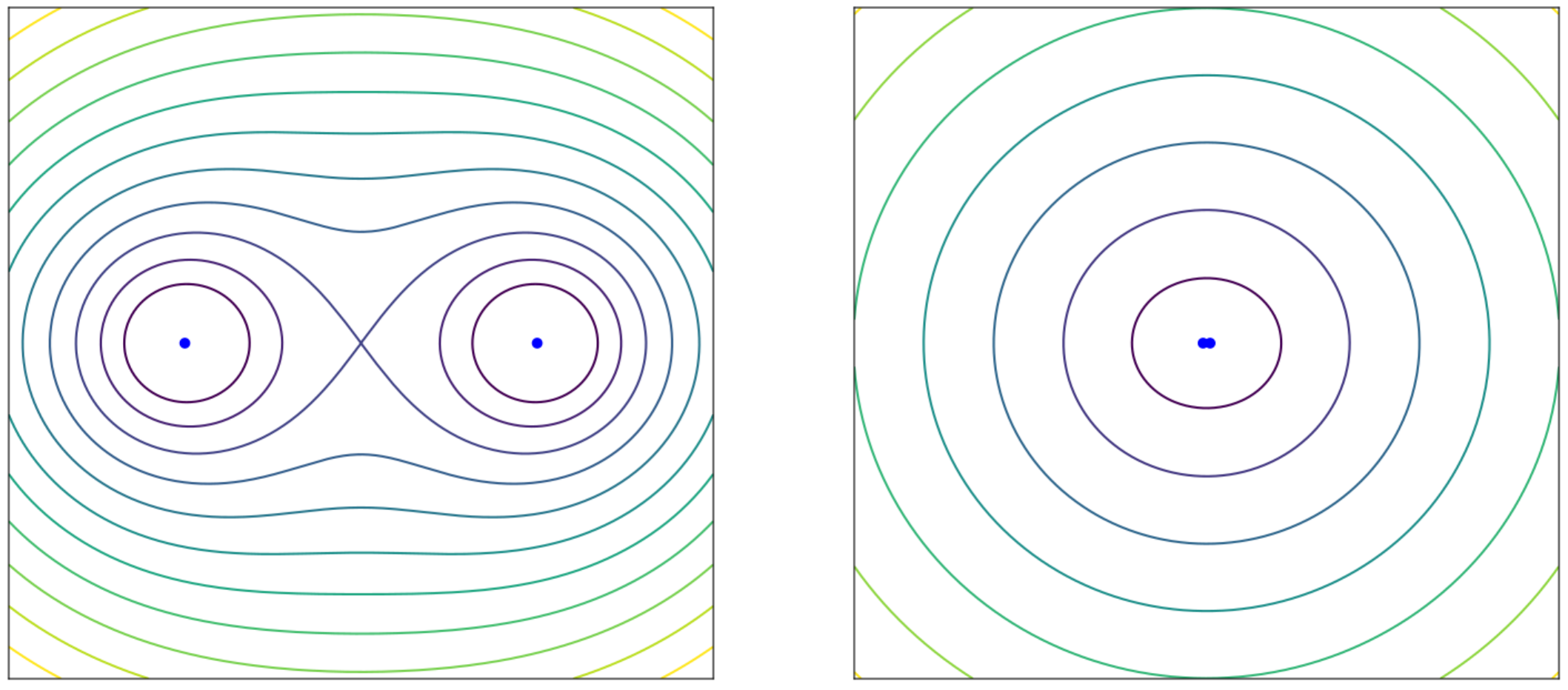}
		\caption{Level sets of the function $\rho$ for $M_{+}=M_{-}=1/2$ and $Z=1$ in Cartesian coordinates $(x,y,z)$. The left figure shows the case where $\rho$ is comparable to the value of $Z$. The right figure shows the case $\rho \gg Z$.  The blue dots represent the two black hole coordinate positions.}
		\label{fig:contourshyp}
	\end{figure}

	For completeness, \Figref{fig:contourshyp} (taken from \cite{Beyer:2018HW}) shows some level sets of the function $\rho$ for small values of $\rho$ on the left and for large values of $\rho$ on the right for $M_+=M_-=1/2$ and $Z=1$. These surfaces undergo a topology change at $\rho=\rho_{crit}$ given by
	\begin{align}
		\rho_{crit} = \frac{\left( M_+ + M_- \right)\left( Z_+ + Z_- \right)}{\left( \sqrt{M_+} + \sqrt{M_-} \right)^{2}}.
		\label{Eq:BirfurificationPoint}
	\end{align}
	Each level set given by $\rho<\rho_{crit}$ is the union of two disconnected $2$-spheres while for $\rho>\rho_{crit}$ it is diffeomorphic to a single $2$-sphere. In this paper we only study initial data sets in the \emph{exterior region} of $\mathbb{R}^3$ given by $\rho>\rho_{crit}$. It is evident from \Figref{fig:contourshyp} that the level sets approach round $2$-spheres for large values of $\rho$ in consistency with \Eqref{Eq:rhoDecay}.
	
	Given the function $\rho$ in \Eqref{Eq:rho_def}, we must now make choices for the function $V$ and the tensor $\dot{\gamma}_{ab}$. For $V$ we pick
	\begin{equation}
		\label{eq:VchoiceBBH}
		V = 1-\frac{27 M^2}{\rho^2},
	\end{equation}    
	where $M=M_{+} + M_{-}$ is the Bondi mass of the background initial data set in consistency with the spherically symmetric single black hole case in \Sectionref{SubSec:Hyperboloidal_Sphereical}. Motivated by the same case (we restrict to $\lambda=0$ in all of what follows), see \Eqsref{GeneralFreeDataNew} and \eqref{Eq:etaDef}, we also set
	\begin{equation}
		\begin{split}
			\delta\kappa=\frac{\rho (2-V)\eta(\rho\, ;M)\partial_{\rho}V + 2 M (1-V)^2+\rho^2 \partial_{\rho}V}{\rho (1-V)^2\eta(\rho\, ;M)},
			\quad
			\delta q=4\frac{\eta(\rho\, ;M)-\rho V}{\rho^2 (V-1)},
			\\
			\delta Q_\Sph {ab} = -\frac{2 V }{({1-V})f}( \kstar{}_\Sph{ab}-\frac{1}{2}\kstar h_\Sph{ab} ),
			\quad
			\delta p_{A}=0.
		\end{split}
		\label{Eq:ydot_Choice}
	\end{equation}    
	Here $f$, $\kstar_\Sph {ab}$ and $\kstar$ are given by the formulas in \Sectionref{SubSec:Kerr_Shild_Like} and \Sectionref{SubSec:Consequences_for_Kerr-Schild-like_data_sets} together with \Eqsref{eq:defkStar} and \eqref{eq:defkStar2}.
	The particular choice of $\delta Q_\Sph {ab}$ here  does clearly not agree with \Eqref{GeneralFreeDataNew} except in the \emph{single black hole case} $M_-=0$ or $Z=0$. The rational for this ``artificial'' choice of $\delta Q_\Sph {ab}$ is that it implies that the resulting background tensor field $Q_{AB}$ identically vanishes, while
	the ``more natural'' choice $\delta Q_{AB}=0$ would in general violate the divergence condition $D_\Sph a Q_\Sph {bc} (\Omega^{-1})^\Sph {ab}=0$ at $\rho=\infty$ of \Propref{res:smoothness}.   In any case, we emphasise that in the \emph{single black hole case} given by $Z=0$ or $M_-=0$, this background initial data set reduces to the one considered in \Sectionref{SubSec:Hyperboloidal_Sphereical} (for $\lambda=0$).
	
	We point out here that $M$ is in general not a `physical' mass as our chosen background does not, in general, satisfy the constraints. Only in the special single black hole case $Z=0$ or $M_-=0$, the quantity $M$ agrees with  Bondi mass.

	We also note that with the above choices, $\kstar$ can be shown to satisfy
	\begin{align}
		\kstar=-\frac{2}{\rho}+\DecayO{\rho}{3}.
		\label{kStar}
	\end{align}
	It is therefore clear that the parabolicity condition \Eqref{eq:parabolcond} holds for all sufficiently large $\rho$. This is consistent with \Propref{res:smoothness}. In numerical calculations we always calculate $\kstar$ in order to verify that the parabolicity condition is satisfied on the \emph{whole} computational domain. One can also demonstrate that for any $M_+,M_-$ and $Z$ as above, the resulting free data fields $(\kappa,B_\Sph{a},h_\Sph{ab},h^\Sph{ab},Q_\Sph{ab})$ satisfy the hypothesis about the free data in \Propref{res:smoothness} (the hypothesis about the unknown fields $A,q$ and $p_\Sph a$ can clearly not be verified a-priori). One of the primary goals of the following subsections is to provide numerical evidence that the unknowns $(A,q,p_\Sph{a})$ and $p_\Sph a$ satisfy (at least some of) the a-priori assumptions of \Propref{res:smoothness}. This gives us confidence that the conclusions of \Propref{res:smoothness}, especially that the resulting vacuum initial data sets are  asymptotically hyperboloidal and extend smoothly to $t=0$, also hold.

	\subsection{Numerical implementation}
	\label{sec:numimpl}
	Given a binary black hole background data set as constructed in \Sectionref{Sec:A_superposition_method_for_generating_binary_black_hole_data}, the next task is to numerically solve the Cauchy problem of \ParabolicHyperbolicR{} with free data (and in some cases also the Cauchy data, see below) determined by this background. While the background data sets are given in Cartesian coordinates $(x,y,z)$ on $\Sigma$, or, equivalently, in corresponding spherical coordinates $(r,\theta,\phi)$ using \Eqref{Eq:r_def}, the $2+1$-decomposition underlying \ParabolicHyperbolicR{} assumes adapted coordinates $(\rho,\vartheta,\varphi)$ where $\rho$ given by \Eqref{Eq:rho_def} labels the leaves of the foliation and  $(\vartheta,\varphi)$ are intrinsic polar coordinates on each $\rho=\mathrm{const}$-surface diffeomorphic to $\mathbb S^2$. As in \cite{Beyer:2018HW, Beyer:2020HW} we choose
	\[\vartheta=\theta,\quad\varphi=\phi\]
	for simplicity.
	This together with \Eqref{Eq:rho_def} therefore completely fixes the transformation between the two coordinate systems $(r,\theta,\phi)$ and $(\rho,\vartheta,\varphi)$ on $\Sigma$ and allows us to write the background data sets in \Sectionref{Sec:A_superposition_method_for_generating_binary_black_hole_data} in the required $(\rho,\vartheta,\varphi)$-coordinates. For further details we refer to \cite{Beyer:2018HW, Beyer:2020HW}.

	Since the computational domain is foliated by $2$-spheres, we can apply the \emph{spin-weight formalism} following  \cite{Penrose:1984tf,Beyer:2015bv,Beyer:2014bu,Beyer:2016fc,Beyer:2017jw,Beyer:2017tu}. A brief summary is given in \Sectionref{Sec:SWSHstuff} in the appendix. We  express the covariant derivative operator $D_\Sph a$ (defined with respect to the intrinsic metric $h_\Sph{ab}$)  in terms of the covariant operator $\hat D_\Sph a$ defined with respect to the round unit-sphere metric $\Omega_\Sph {ab}$; recall that $D_\Sph a-\hat D_\Sph a$ can be expressed in terms of smooth intrinsic tensor fields. Using \Sectionref{Sec:SWSHstuff}, we can then express the covariant derivative operator $\hat D_\Sph a$ in terms of the $\eth$- and $\eth'$-operators. Once all of this has been completed for all terms in \ParabolicHyperbolicR, each term of each of these equations  ends up with a consistent well-defined spin-weight. Most importantly, however, all terms are explicitly regular: Standard polar coordinate issues at the poles of the $2$-sphere disappear when all quantities are  expanded in terms of \keyword{spin-weighted spherical harmonics} and \Eqsref{eq:eths} and \eqref{eq:eths2} are used to calculate the intrinsic derivatives. From the numerical point of view this gives rise to a (pseudo)-spectral scheme. We can  largely reuse the code presented in \cite{Beyer:2018HW,Beyer:2020HW} subject to three minor changes: (1) the definition of $\rho$ now allows that $Z_{+}\ne Z_{-}$ in agreement with \Eqref{Eq:Z_choices}, (2) the definition of $V$ is changed agreement with \Eqref{eq:VchoiceBBH}, and, (3) the definition of $\dot{\gamma}_{ab}$ is changed in agreement with \Eqref{Eq:ydot_Choice}. These three changes do not significantly affect our numerical methods. 
	Once these changes had been made to the code, convergence tests (analogous to the ones presented in \cite{Beyer:2018HW}) were carried out successfully. All of the following simulations were carried out using the adaptive SciPy ODE solver \textit{{odeint}}\footnote{See \url{https://docs.scipy.org/doc/scipy/reference/generated/scipy.integrate.odeint.html}.}.

	Notice that the background data sets constructed in \Sectionref{Sec:A_superposition_method_for_generating_binary_black_hole_data} are axially symmetric and hence there is no dependence on the angular coordinate $\varphi=\phi$. Motivated by this we restrict to numerical solutions of \ParabolicHyperbolicR{} with that same symmetry in all of what follows. We can therefore restrict to the axisymmetric case of the spin-weight formalism in \Sectionref{Sec:SWSHstuff}.

	\subsection{Axisymmetric perturbations of single Schwarzschild black hole initial data}
	\label{SubSec:Perturbing_spherically_symmetric_initial_data_sets}
	
	In this section now, we use the background data set given in \Sectionref{Sec:A_superposition_method_for_generating_binary_black_hole_data} with the choices $M_{+}=1$, $M_{-}=Z=0$ so that the background initial data set reduces to the spherically symmetric single black case first introduced in \Sectionref{SubSec:Hyperboloidal_Sphereical} with $M=1$ (for $\lambda=0$). This background and therefore the free data for \ParabolicHyperbolicR{}  given by \Eqref{GeneralFreeDataNew} with $V = 1 - 27M^2\rho^{-2}$ are therefore spherically symmetric. The fields
	\begin{align}
		\mathring{q} =\frac{2 \sqrt{3} (1-3 \rho^{-1}) \sqrt{1+6 \rho^{-1}}}{9}, \;\;\mathring{A}=\frac{\sqrt{27}}{\rho},\;\;\mathring{p}_{a}=0,
		\label{BackgroundSolution}
	\end{align}
	agree with the particular solution of \ParabolicHyperbolicR{} given by \Eqref{GeneralSolNew} (or by \Eqsref{eq:sphsolq} and \eqref{eq:sphsolA} for $\C=0$ and $m=1$) representing single unperturbed spherically symmetric Schwarzschild black hole initial data of unit mass. The purpose of the present subsection is to generate axisymmetric (non-linear) \emph{perturbations} of  this solution by solving \ParabolicHyperbolicR{} with the same free data, but with the following \emph{perturbed Cauchy data} imposed at\footnote{For the single black-hole case, we have $\rho_{crit}=0$, see \Eqref{Eq:BirfurificationPoint}.} the initial radius $\rho_{0}=5$:
	\begin{align}
		\label{eq:perturbedCD}
		\left. q \right|_{\rho=\rho_{0}}= \left. \mathring{q} \right|_{\rho=\rho_{0}}+\epsilon\sin\left( \theta \right),\;\; \left. A \right|_{\rho=\rho_{0}}= \left. \mathring{A} \right|_{\rho=\rho_{0}}+\epsilon\sin\left( \theta \right),\;\; \left. p_\Sph a \right|_{\rho=\rho_{0}}= 0,
	\end{align}
	for some arbitrary constant $\epsilon\in\mathbb R$. Especially for small values of $\epsilon$, we can interpret the resulting vacuum initial data sets as (nonlinear) {perturbations} of single Schwarzschild black hole initial data.

	We express \ParabolicHyperbolicR{} for these free data and Cauchy data numerically in terms of the formalism in \Sectionref{Sec:SWSHstuff}:
	\begin{align}
		\partial_{\rho}A &=  -\frac{\rho}{2}\left( \kappa q + \frac{1}{4}q^2  \right)A^3 +\frac{1}{2 \rho}\left( \left( 1 + A\eth\left( {\eth^\prime}\left( A \right) \right) \right)A - \left( 1- 2p\pbar \right) \right),
		\label{Org_GeneralPDE_dA_2}
		\\
		\partial_{\rho}q &= \frac{1}{\sqrt{2}\rho^2}\left( {\eth^\prime}\left( p\right) + \eth\left( \pbar \right)  \right)A -\frac{1}{\rho}\left( q - 2\kappa \right)+\frac{\sqrt{2}}{\rho^2}\left( p{\eth^\prime}\left(A \right)+\pbar\eth\left(A\right) \right),
		\label{Org_GeneralPDE_dq_2}
		\\
		\partial_{\rho}p &=\frac{A}{2\sqrt{2}}\eth\left(q\right) -\frac{2}{\rho}p +\frac{1}{\sqrt{2}}\left( \kappa -\frac{1}{2}q \right)\eth\left( A \right) ,
		\label{Org_GeneralPDE_p2_2}
		\\
		\partial_{\rho}\pbar &= \frac{A}{2\sqrt{2}}{\eth^\prime}\left(q\right) -\frac{2}{\rho}\pbar +\frac{1}{\sqrt{2}}\left( \kappa -\frac{1}{2}q \right){\eth^\prime}\left( A \right),
		\label{Org_GeneralPDE_p1_2}
	\end{align}
	where
	\begin{align}
		\label{Eq:pDef}
		p=\frac 1{\sqrt 2} p_\Sph a \left(\partial_{\vartheta}^\Sph a-{\text{i}}\, {\csc\theta}\, \partial_\varphi^\Sph a\right),\quad
		\bar p=\frac 1{\sqrt 2} p_\Sph a \left(\partial_{\vartheta}^\Sph a+{\text{i}}\, {\csc\theta}\, \partial_\varphi^\Sph a\right),
	\end{align}
	and, see \Eqref{eq:sphsolkappa} for $M=1$,
	\begin{align}
		\kappa=\frac{\rho^{2} +3\rho+9}{\rho^{3/2}\sqrt{ 27(\rho+6)}}.
	\end{align}
	The quantities $A$ and $q$ have spin-weight zero, while $p$ and $\pbar$ have spin-weight $1$ and $-1$, respectively. Our symmetry assumptions (and our particular representation of the bundle of orthonormal frames on $\mathbb S^2$) allows us to assume that
	\[p=\pbar.\]

	Let us present the numerical results now. To this end we define the $\sup$-norm over $\mathbb S^2$ for any smooth scalar function $\mathcal F(\rho,\vartheta)$ (such as $A$ and $q$ above) as
	\begin{align}
		\| \mathcal{F} \|(\rho)=\max_{\vartheta\in [0,\pi]}| \mathcal{F}(\rho,\vartheta) |,
		\label{Eq:SupNorm}
	\end{align}
	while for the covector $p_\Sph a$, this norm is defined as
	\begin{align}
		\| p \|(\rho)=\max_{\vartheta\in [0,\pi]}\sqrt{\left(\Omega^{-1}\right)^\Sph{ab}p_\Sph a (\rho,\vartheta) p_\Sph b (\rho,\vartheta)}
		=\max_{\vartheta\in [0,\pi]}\sqrt{p (\rho,\vartheta)\pbar (\rho,\vartheta)}.
	\end{align}

	\begin{figure}[t]		
		\centering
		\includegraphics[width=0.385\linewidth]{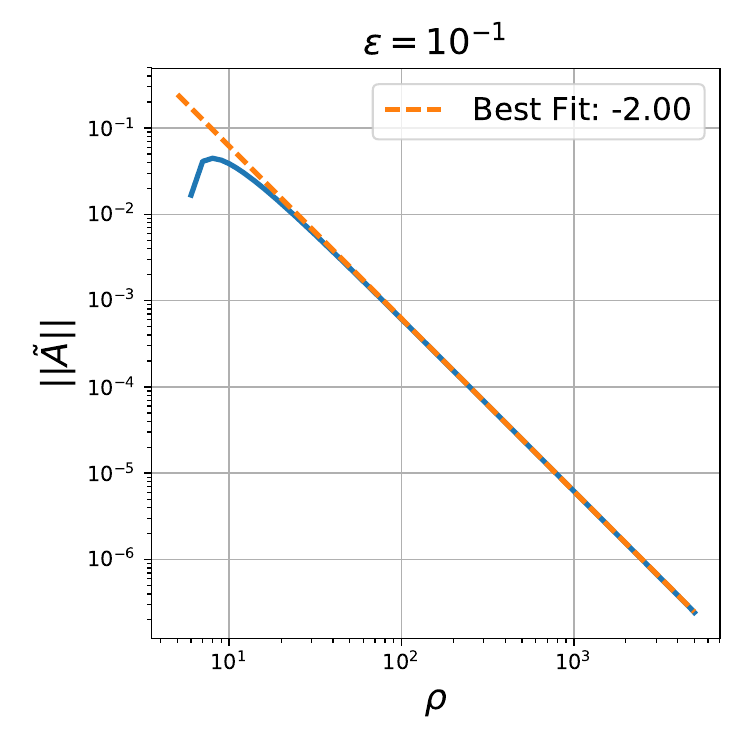}			
		\includegraphics[width=0.385\linewidth]{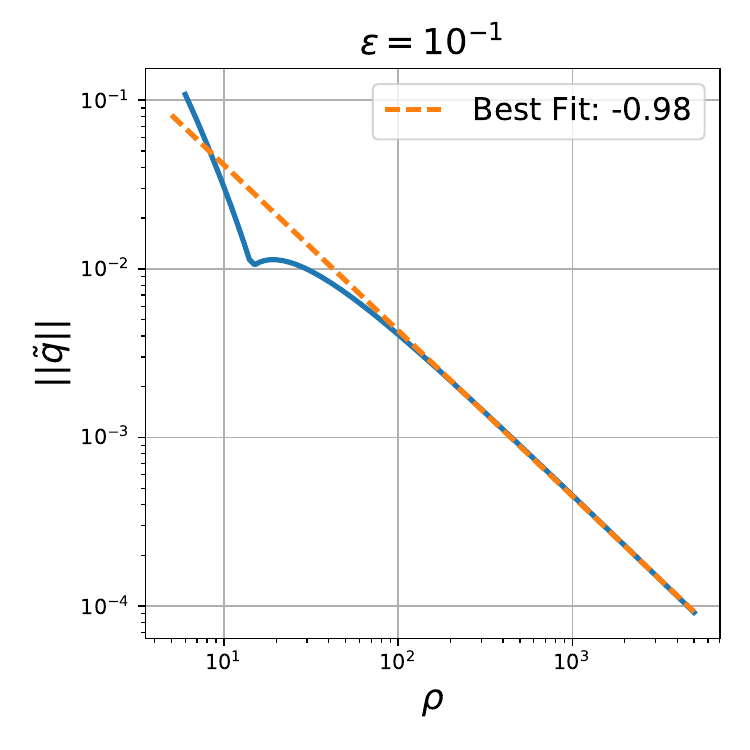}
		\\
		\includegraphics[width=0.385\linewidth]{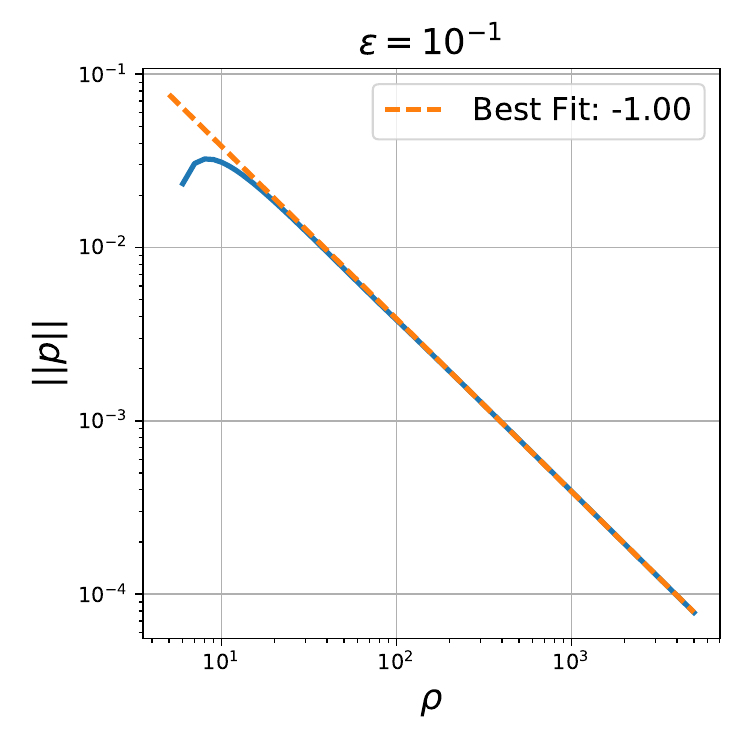}			                    
		\caption{The leading order asymptotics for the ``single black hole case" obtained with $\epsilon=10^{-1}$, $N=11$ and error tolerance of $10^{-10}$. The definitions of $\tilde q$ and $\tilde A$ are given in \Eqref{eq:defTilde}. Each numerical curve (solid blue) in the three plots is fitted to the function $C \rho^k$ (dashed yellow)  for some $C>0$ where ``Best fit'' gives the best value for $k$.}
		\label{fig:phypsbh1}
	\end{figure}
	
	To discuss the expected behaviour we now introduce the following quantities
	\begin{equation}
		\label{eq:defTilde}
		\tilde{q}=q-\frac{2}{\sqrt{27}M},
		\quad
		\tilde{A}=A -\frac{\sqrt{27}M}{\rho}.
	\end{equation}    
	According to \Propref{Result:AsymAnal_Hyp_R} together with \Sectionref{SubSec:Consequences_for_Kerr-Schild-like_data_sets}, we expect the following behaviour
	\begin{align}
		\label{eq:Result2Exp}
		\left\| \tilde A \right\|=\DecayO{\rho}{2},
		\quad
		\left\| \tilde q\right\|=\DecayO{\rho}{},
		\quad
		\|p \| = \DecayO{\rho}{}.
	\end{align} 
	This is confirmed by the first three plots of \Figref{fig:phypsbh1} for  $\epsilon=10^{-1}$, an absolute and relative error tolerance for the adaptive ODE solver of $10^{-10}$, and for $N = 11$, where $N$ is the number of spatial points in the $\vartheta$-direction (recall that due to axisymmetry, there is no $\varphi$-dependence). The plots shown in \Figref{fig:phypsbh1} show that our numerically constructed solutions are asymptotic hyperboloidal (at least with minimal regularity at infinity) and are compatible with \Propref{Result:AsymAnal_Hyp_R}.
	
	\begin{figure}[t!]		
		\centering
		\includegraphics[width=0.4\linewidth]{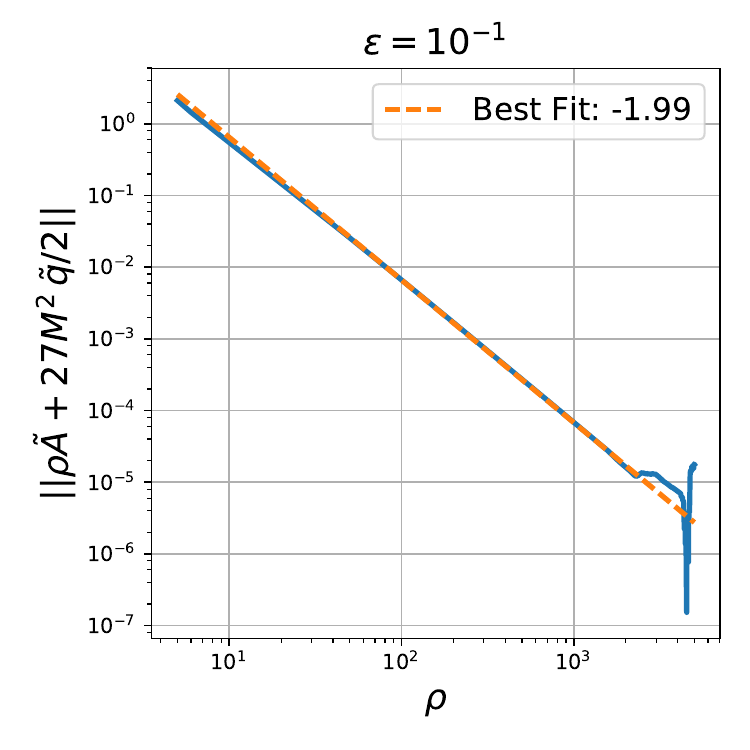}			
		\includegraphics[width=0.4\linewidth]{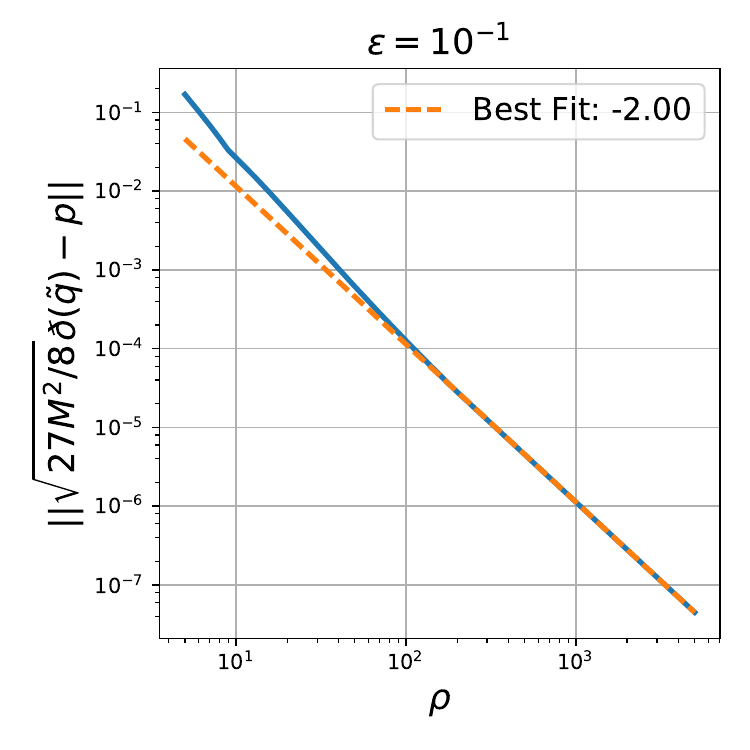}
		\\
		\includegraphics[width=0.4\linewidth]{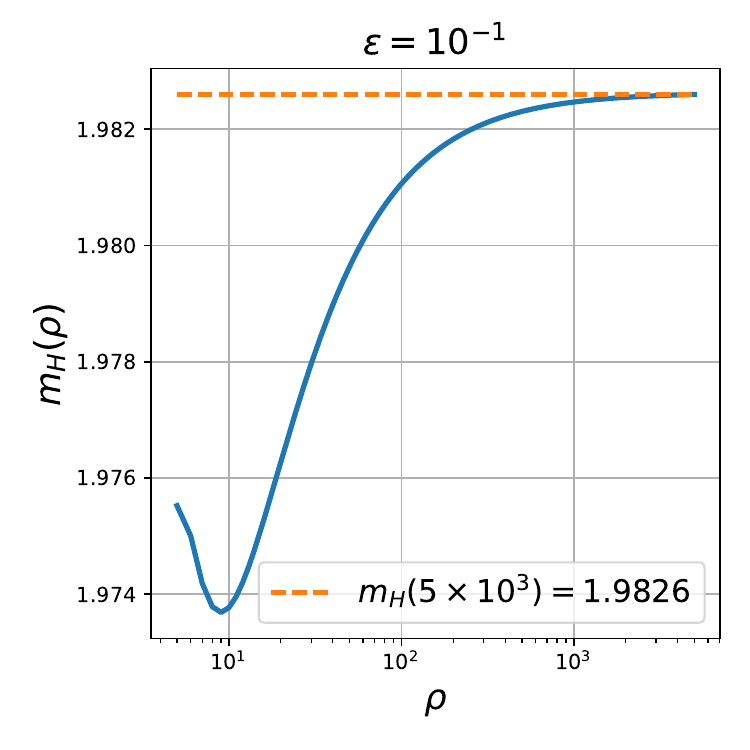}	
		\caption{Higher order asymptotics and evolution of the Hawking mass $m_H$ for the ``single black hole case" obtained with the same parameter values as in \Figref{fig:phypsbh1}.}
		\label{fig:decays_2_ssb}
	\end{figure}
	Given that our background fields satisfy the assumptions of \Propref{res:smoothness} we may now wonder to what extent it is possible to also verify the asymptotic expansions \Eqsref{eq:AHExpCOeffFirst}--\eqref{eq:AHExpCOeffLast} given by that result. To this end we note that \Propref{res:smoothness} (together with the formulas in \Sectionref{SubSec:Kerr_Shild_Like}) yields that
	\begin{align*}
		\tilde{q}=\frac{q^{(1)}}{\rho}+\DecayO{\rho}{2},
		\quad
		\tilde{A}=-\frac{27M^2}{2\rho^{2}}q^{(1)}+\DecayO{\rho}{3},
		\quad
		p = \sqrt{\frac{27M^2}{8}}\frac{\eth(q^{(1)})}{\rho} + \DecayO{\rho}{2}.
	\end{align*}
	Combining these, we find that
	\begin{align*}
		\sqrt{\frac{27M^2}{8}}\eth(\tilde{q})-p=\DecayO{\rho}{2},
		\quad
		\rho\tilde{A} + \frac{27M^2}{2}\tilde{q}=\DecayO{\rho}{2},
	\end{align*}
	and hence we have
	\begin{align}
		\| \sqrt{27M^2/8}\,\eth(\tilde{q})-p \|=\DecayO{\rho}{2},
		\quad
		\| \rho\tilde{A} + {27M^2}\tilde{q}/2 \|=\DecayO{\rho}{2}.
		\label{Eq:Result3_Decays}
	\end{align}
	These theoretically obtained asymptotics are indeed confirmed in the first two plots in \Figref{fig:decays_2_ssb}. Together, \Figsref{fig:phypsbh1} and \ref{fig:decays_2_ssb} verify our theoretical results for $q^{(0)}$, $q^{(1)}$, $A^{(0)}$, $A^{(1)}$, $A^{(2)}$, $p_{B}^{(0)}$ and $p_{B}^{(1)}$. However, the numerical results do not seem to be accurate enough to construct $q^{(2)}$ or $A^{(3)}$.
	
	As mentioned earlier, it turns out that the direct numerical estimation of the Bondi mass via \Eqsref{eq:hatMBondiMass1} and \eqref{eq:hatMBondiMass2} is unsuccessful. The limit appears to diverge as a consequence of numerical errors. This is why we determine Bondi mass by evolving the quantity $m_H$ by means of \Eqsref{eq:Mevol} and \eqref{eq:defFa} simultaneously with \Eqsref{Org_GeneralPDE_dA_2}--\eqref{Org_GeneralPDE_p1_2}. Since the background initial data set is spherically symmetric, so in particular $\kstar=-2/\rho$, the evolution equation for $m_{H}$ takes the form here        
	\begin{align}
		\partial_{\rho} m_H= \frac{\rho^2}{8}\underline{F_{[a]}},
		\label{Eq:HawkingMass_SS}
	\end{align}
	where we use $\kstar=-2/\rho$, $t=1/\rho$, \Eqref{eq:average},  and
	\begin{equation}
		\begin{split}
			\underline{F_{[a]}}= \int_{d\Omega} { \frac{8}{\rho^2}p\pbar } +{\sqrt{2}}q^{2}\left( \bar{p}\, \eth\left( \frac{A}{q} \right) + {p}\, \eth^\prime\left( \frac{A}{q} \right) \right)\rho -\frac{4}{\rho^2}A^{-2}\eth(A){\eth^\prime}(A) \,d\Omega.
		\end{split}
	\end{equation}
	The numerically calculated function $m_H(\rho)$ is shown in the last plot of \Figref{fig:decays_2_ssb}. Here we see that it quickly converges to a constant number approximating the Bondi mass $m$ as $m = m_{H}(\rho)$ for $\rho=5\times 10^{3}$. It is of course natural to wonder how good this approximation is. For this we consider the quantity 
	\begin{align}
		\mathcal{E}_{A}[m]=| m_{H}(2\rho)- m_{H}(\rho)| ,
	\end{align}
	which is calculated for $\rho=5\times 10^{3}$, as an approximation of the absolute (and approximately also relative) error. For our example case with $\epsilon=10^{-1}$ we find
	\begin{align}
		m = 1.9826,
		\quad
		\mathcal{E}_{A}[m]=1.654\times 10^{-6}.
	\end{align}
	Notice that the main error source here is likely the error associated with measuring $m$ at a finite value of $\rho$. However, due to the errors generated by numerically solving the constraints for very large values of $\rho$, we find that evaluating the mass at $\rho=5\times 10^{3}$ is optimal. Indeed, in the first plot of \Figref{fig:decays_2_ssb} we see that the numerical errors become important at around $\rho\sim 10^3$.

	\subsection{Binary black hole-like initial data sets}

	\begin{figure}[t]		
		\centering
		\includegraphics[width=0.385\linewidth]{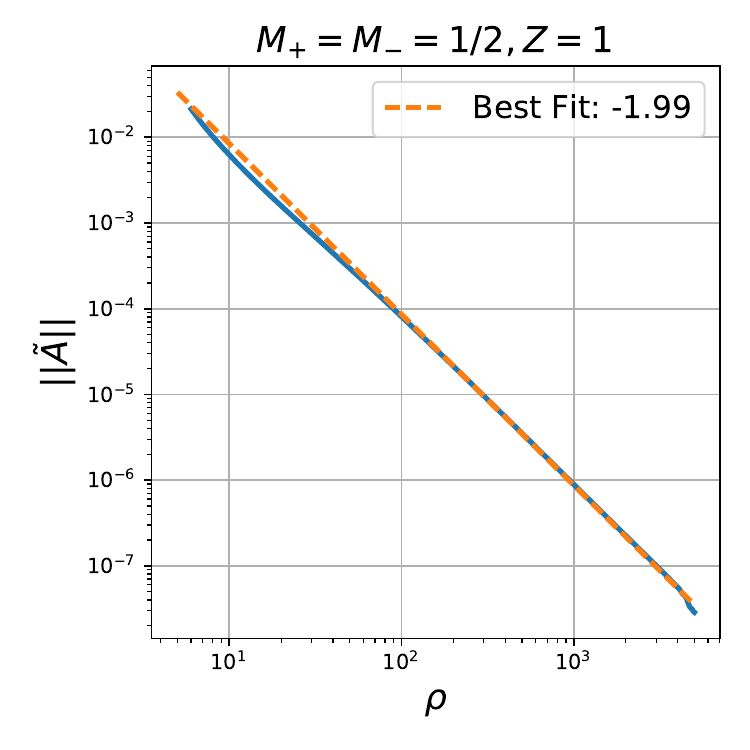}			
		\includegraphics[width=0.385\linewidth]{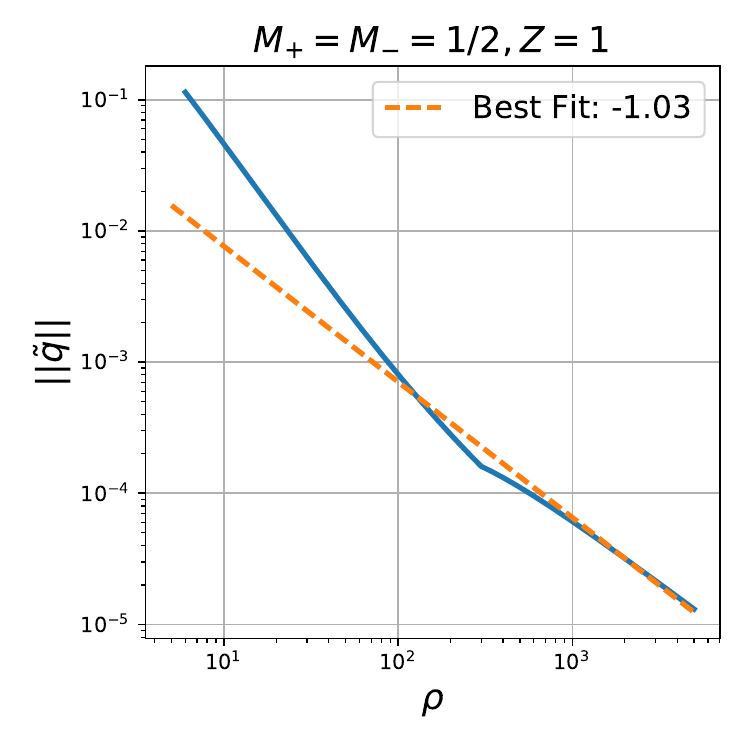}
		\\
		\includegraphics[width=0.385\linewidth]{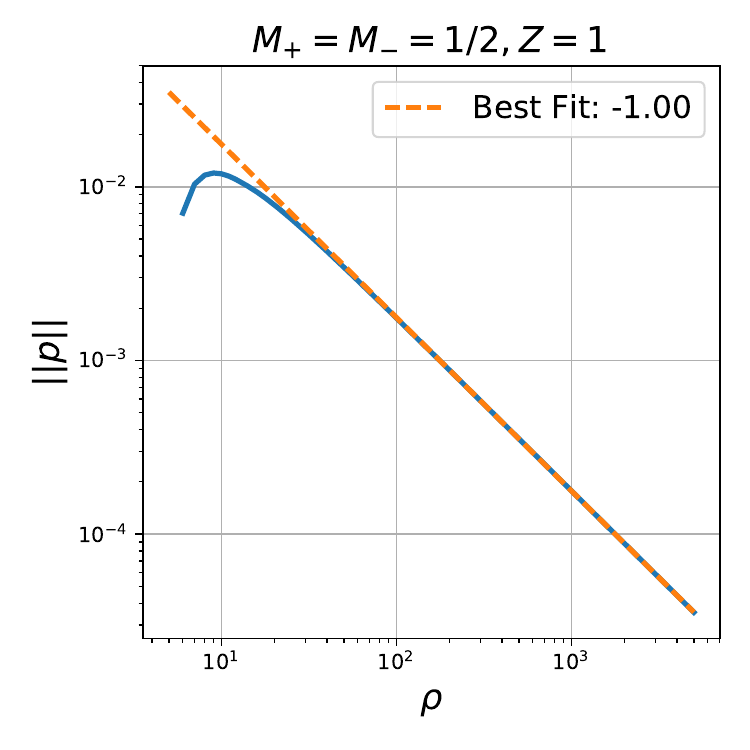}       
		\caption{The leading order asymptotics for the ``binary black hole case" obtained with $M_+=M_-=1/2$, $\rho_0 = 3$, $N = 11$ and numerical error tolerance of $10^{-10}$. Each numerical curve (solid blue) in the three plots is fitted to the function $C \rho^k$ (dashed yellow)  for some $C>0$ where ``Best fit'' gives the best value for $k$.}
		\label{fig:mhypbbh1}
	\end{figure}

\begin{figure}[t]	
		\centering
		\includegraphics[width=0.4\linewidth]{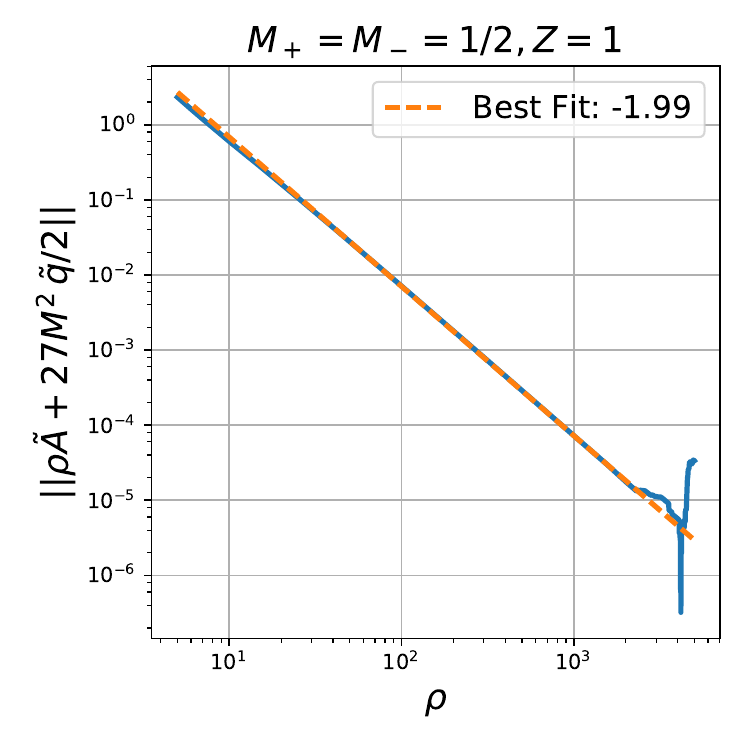}			
		\includegraphics[width=0.4\linewidth]{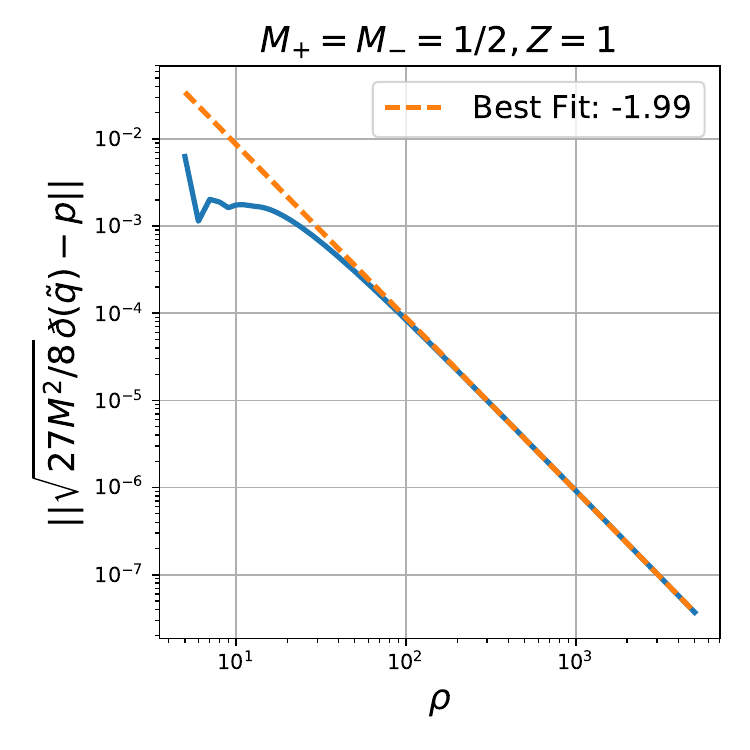}	
		\\
		\includegraphics[width=0.4\linewidth]{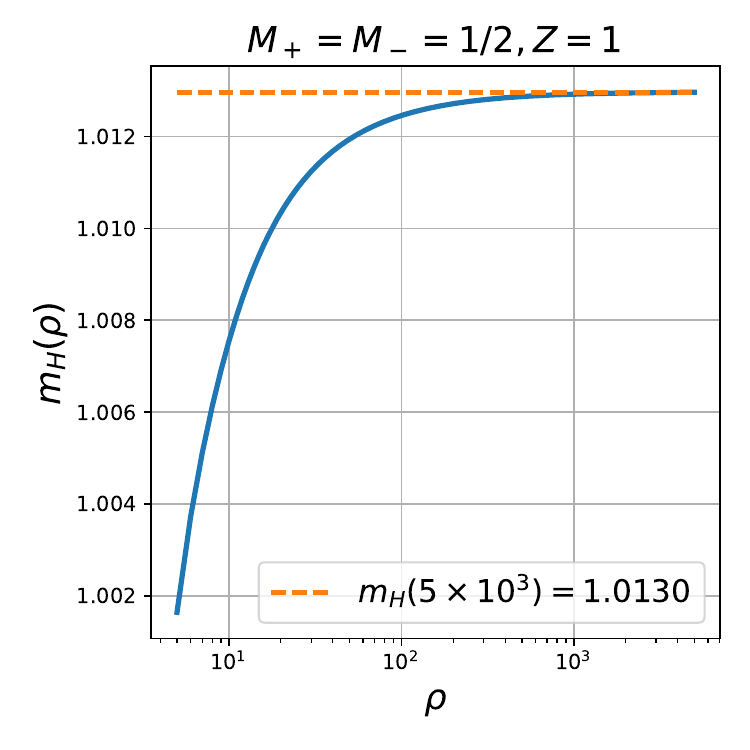}
		\includegraphics[width=0.4\linewidth]{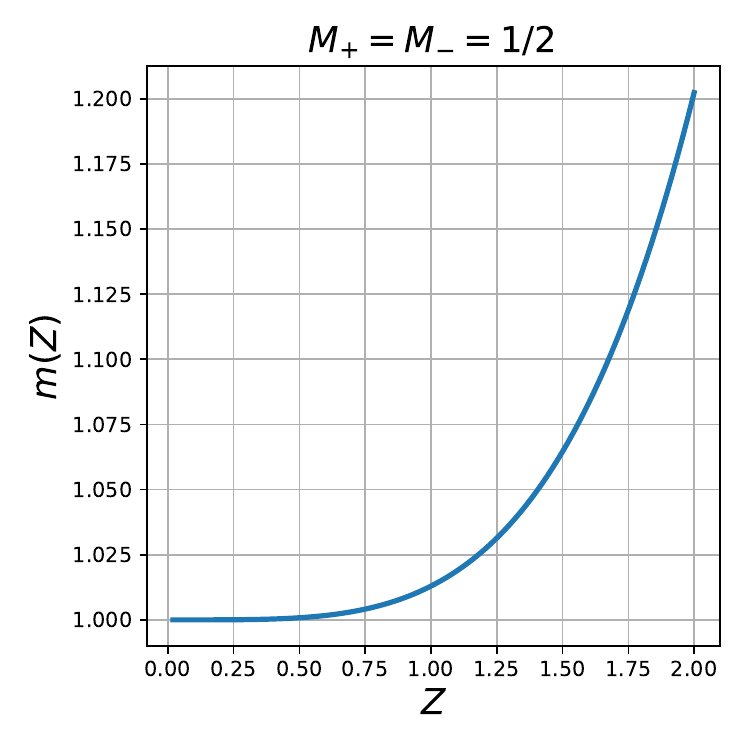}         
		\caption{Higher order asymptotics and evolution of the Hawking mass $m_H$ for the ``binary black hole case" obtained with the same parameter values as in \Figref{fig:mhypbbh1}.}
		\label{fig:mofzbbhhyp}
	\end{figure}

	In this subsection we  essentially repeat the same numerical experiments as before with two changes: (1), the background initial data set is now determined with parameters $M_{+}=M_{-}=1/2$ and $Z=1$ (an ``equal mass binary black hole case''), and (2),  instead of the ``perturbed'' Cauchy data as in \Eqref{eq:perturbedCD}, we now choose the values obtained from the background data set at $\rho_{ 0 }=5$. For this particular case \Eqref{Eq:BirfurificationPoint} gives  $\rho_{crit}=1$. Our numerical findings, as shown in the first three plots of \Figref{fig:mhypbbh1} are again consistent with \Eqref{eq:Result2Exp} as expected from \Propref{Result:AsymAnal_Hyp_R}. Similarly, the first two plots of \Figref{fig:mofzbbhhyp} are consistent with \Eqref{Eq:Result3_Decays} and therefore, as with the ``single black hole case", support \Propref{res:smoothness}. We interpret this as strong evidence that the vacuum initial data sets we have numerically calculated are asymptotically hyperboloidal.
	
	Constructing the Bondi mass in the same way as in \Sectionref{SubSec:Perturbing_spherically_symmetric_initial_data_sets} yields the third plot in \Figref{fig:mofzbbhhyp}, and
	\begin{align}
		m = 1.01297,
		\quad
		\mathcal{E}_{A}[m]=3.78\times 10^{-6}.
	\end{align}

	For fixed values of $M_+$ and $M_-$, say, $M_+=M_-=1/2$ as before, one expects the resulting Bondi masses to depend strongly on the separation parameter $Z$. To investigate this we numerically calculate the resulting vacuum initial data sets and Bondi masses for a range of separation parameter values $Z$. Since we treat $\rho_{ 0 }=5$ as fixed, \Eqref{Eq:BirfurificationPoint} introduces an upper bound for the possible values for $Z$, namely $Z<\rho_{0}$. The results of our numerical calculations are shown in the last plot of \Figref{fig:mofzbbhhyp}. Observe that the Bondi mass turns out to be an increasing function of  $Z$. For $Z=0$ (in which the single black hole solution is obtained) we find that $m=1$, as expected. Observe that the case $Z=0$ here is different from the single black hole case in \Sectionref{SubSec:Perturbing_spherically_symmetric_initial_data_sets} because the Cauchy data are determined differently. When $Z$ is now increased we find that the Bondi mass becomes larger. This is intuitive as one expects the (negative) gravitational binding energy to become small as the separation distance $Z$ is increased. However, it is interesting to compare this to our results in \cite{Beyer:2018HW, Beyer:2020HW} for asymptotically flat initial data sets, where we found that the ADM mass \emph{decreases} as a function of the separation distance $Z$.   
	
	\section{Conclusions}
	In this paper we discuss the asymptotic behaviour of vacuum initial data sets constructed as solutions of a parabolic-hyperbolic formulation of the vacuum constraint equations.
	The primary goal of this work is to establish whether or not it is possible to reliably construct asymptotically hyperboloidal initial data sets using R{\'a}cz's parabolic-hyperbolic formalism.	
	
	We found that initial data sets constructed as solutions of R{\'a}cz's original parabolic-hyperbolic formulation of the constraints on  asymptotically hyperboloidal background initial data sets are, in gen\-eral, asymptotically hyperboloidal with a well-defined Bondi mass. In particular, no modifications are necessary (in contrast to the asympotically flat case) as long as the background initial data set used to construct the free data for the equations satisfy certain properties at $\rho=\infty$. In fact we provide explicit conditions together with strong evidence guaranteeing that $\log\rho$-terms in expansions at $\rho=\infty$ are completely ruled out and the resulting vacuum initial data sets therefore extend to infinity smoothly. This contrasts the more traditional conformal approach where it was found \cite{PhD:HypDef} that asymptotically hyperboloidal initial data sets are in general poly-logarithmic at infinity unless the condition $K=const$ in \cite{PhD:HypDef} holds.

	%\newpage
	
	\section*{Acknowledgements}
	
	JR was supported by a Ph.D scholarship awarded by the University of Otago. 
	
	\begin{appendices}
		\setcounter{equation}{0} % To restart the counting 
		\numberwithin{equation}{section} % To counting in the appendix 
		\appendixpage
		\appendix
		
		\section{Spin-weight and spin-weighted spherical harmonics}\label{Sec:SWSHstuff}
		We say that a function $f$ defined on $\mathbb{S}^2$ has \emph{spin-weight $s$} if it transforms as $f \to e^{\text{i} s \xi} f$ under a local rotation by an angle $\xi$ in the tangent plane at any point in $\mathbb{S}^2$. Let $(\vartheta,\varphi)$ be standard polar coordinates on $\mathbb S^2$. If $f$ has spin-weight $s$ and is sufficiently smooth, it can be written as
		\begin{equation}\label{eq:functionS2}
			f(\vartheta,\varphi)=  \sum\limits_{l=|s|}^{\infty}  \sum\limits_{m=-l}^{l} f_{lm}\, {}_{s}Y_{lm} (\vartheta,\varphi),
		\end{equation}
		where $_{s}Y_{lm}( \vartheta , \varphi)$ are the \emph{spin-weighted spherical harmonics (SWSH)} and where $f_{lm}$ are complex numbers. Using the conventions in \cite{Penrose:1984tf,Beyer:2015bv,Beyer:2014bu,Beyer:2016fc,Beyer:2017jw,Beyer:2017tu}, these functions satisfy
		\begin{equation}\label{integral_properties_spherical_harmonics}
			\int \limits_{\mathbb{S}^2} \  {}_{s} Y_{l_1 m_1 }(\vartheta,\varphi)
			\: _{s}\overline{Y}_{l_2 m_2}(\vartheta,\varphi) \ d\Omega = \delta_{l_1 l_2} \delta_{m_1 m_2},
		\end{equation}
		where $\delta_{lm}$ is the Kronecker delta and $d\Omega$ is the area element of the metric of the round unit sphere. Using this we find that the coefficients $f_{lm}$ in \Eqref{eq:functionS2} can be calculated as
		\begin{equation}
			f_{lm}=\int \limits_{\mathbb{S}^2} f(\vartheta,\varphi)\, {}_{s}\overline{Y}_{lm} (\vartheta,\varphi) d\Omega.
		\end{equation}
		
		The \textit{eth-operators} $\eth$ and $\eth'$ are  defined by 
		\begin{equation}\label{eq:def_eths}
			\eth f       = \partial_\vartheta f - \dfrac{\text{i}}{ \sin \vartheta} \partial_\varphi f- s f \cot \vartheta, \quad 
			\eth' f = \partial_\vartheta f + \dfrac{\text{i}}{ \sin \vartheta} \partial_\varphi f + s f \cot \vartheta  ,
		\end{equation}
		for any function $f$ on $\mathbb{S}^2$ with spin-weight $s$. We have
		\begin{align}\label{eq:eths}
			\eth  \hspace{0.1cm}_{s}Y_{lm} (\vartheta,\varphi)  &= - \sqrt{ (l-s)(l+s+1) } \hspace{0.1cm}_{s+1}Y_{lm} (\vartheta,\varphi) , \\
			\label{eq:eths2}
			\eth'   \hspace{0.1cm}_{s}Y_{lm} (\vartheta,\varphi)   &= \sqrt{ (l+s)(l-s+1) } \hspace{0.1cm}_{s-1}Y_{lm} (\vartheta,\varphi) , \\
			\eth' \eth  \hspace{0.1cm}_{s}Y_{lm} (\vartheta,\varphi)   &= - (l-s)(l+s+1) \hspace{0.1cm}_{s}Y_{lm} (\vartheta,\varphi) .
		\end{align}
		Thus, using the properties above it is easy to see that $\eth$ raises the spin-weight by one while $\eth'$ lowers it by one.
		
		In our discussion we are often interested in the \emph{average} of a function $f$ with spin-weight $0$ on $\mathbb{S}^2$ defined by 
		\begin{equation}
			\label{eq:average}
			\underline{f} = \dfrac{1}{4 \pi} \int \limits_{\mathbb{S}^2} \: f d\Omega.
		\end{equation}
		Expressing $f$ in terms of SWSH and using \Eqref{integral_properties_spherical_harmonics} it follows 
		\begin{equation}\label{ec:mean_value_s2}
			\begin{aligned}
				\underline{f} &= \dfrac{1}{4 \pi}  \int \limits_{\mathbb{S}^2} \: \sum\limits_{l=0}^{\infty}  \sum\limits_{m=-l}^{l} f_{lm}\, {}_{0}Y_{lm} (\vartheta,\varphi) \; d\Omega , \\
				&= \dfrac{\sqrt{4\pi}}{4 \pi }  \int \limits_{\mathbb{S}^2} \: \sum\limits_{l=0}^{\infty}  \sum\limits_{m=-l}^{l} f_{lm}\,{}_{0}Y_{lm} (\vartheta,\varphi) \; _{0}\overline{Y}_{00}(\vartheta,\varphi) \; d\Omega , \\
				& =   \frac{1}{\sqrt{4\pi}}f_{00},
			\end{aligned}
		\end{equation}
		where we have used the fact that $_{0}Y_{00}(\vartheta,\varphi) =  (4\pi)^{-1/2}$. 
		Another quantity of interest is the $L^2$-norm with respect to the standard round metric on $S^2$. The \emph{Parseval identity} states that
		\begin{equation}
			\label{eq:L2}
			\|f\|^2_{L^2(\mathbb{S}^2)}=\sum_{l=0}^\infty\sum_{m=-l}^l |f_{lm}|^2.
		\end{equation}
		
		Finally we notice that all functions considered in this paper are axially symmetric and therefore do not depend on the angle $\varphi$. For such functions, all coefficients with $f_{lm}$ with $m\not=0$ vanish and we use the following short-hand notation to write \Eqref{eq:functionS2} as
		\begin{equation}\label{eq:functionS2axial}
			f(\vartheta)=  \sum\limits_{l=|s|}^{\infty}  f_{l}\, {}_{s}Y_{l} (\vartheta).
		\end{equation}

	\section{Asymptotically hyperbolodial initial data sets from other evolutionary formulations of the constraints}
	Although this paper focuses on R\'acz's ``original'' parabolic-hyperbolic system \ParabolicHyperbolicR{}, we would also like present some brief results about other evolutionary formulations of the constraints. The goal of this Appendix here is to introduce and discuss two other evolutionary formulations of the constraints.
	
	\subsection{Other evolutionary formulations of the constraints}
	\label{Sec:Other_evolutionary_formulations_of_the_constraints}	
	\subsubsection{A modified parabolic-hyperbolic formulation of the constraints}
	\label{Sec:A_modified_parabolic_hyperbolic_formulation_of_the_constraints}
	We start with our ``modified'' parabolic-hyperbolic formulation of the vacuum constraints which was first presented by us in \cite{Beyer:2020HW} when considering asymptotically flat initial data sets. 
	
	First, recall that $\kappa$ is one of the free data in the formulation introduced in \Sectionref{SubSec:Racz2Plus1Decomp} while $q$ is one of the unknowns. We modify \ParabolicHyperbolicR{} by introducing a new free data field $\R$ such that
	\begin{equation}
		\label{eq:alternativedatadec}
		\kappa=\R q
	\end{equation}
	where $q$ continues to be an \emph{unknown}. The equations obtained from \ParabolicHyperbolicR{} by replacing all instances of $\kappa$ with $\R q$ are
	\begin{align}
		\overset{\star}{k}\mathcal{L}_{\rho}A+A^2 D^ \Sph{c}D_\Sph{c} A-\overset{\star}{k}B^\Sph{C} D_ \Sph{c}A=\,
		&\frac 12 A^3 E
		+\frac 12A F,
		\label{ParabolicEquation}
		\\
		\mathcal{L}_{\rho}q-B^\Sph{c} D_\Sph{c}q-AD_\Sph{c}p^\Sph{c}-2p^\Sph{c}D_\Sph{c}A=\,&\overset{\star}{k} {}^\Sph{cs}Q_\Sph{cs}+\frac{1}{2}q\overset{\star}{k} -\overset{\star}{k}\R q,
		\label{FinalSystemDiffNorm}
	\end{align}
	\begin{align}
		\begin{split}
			\mathcal L_{\rho} p_\Sph{c} - B^\Sph{s} D_\Sph{s} p_\Sph{c} -A\left( \frac{1}{2}+\R \right)D_\Sph{c}q=\,&p_\Sph{s} D_\Sph{c} B^\Sph{S}
			-AD_\Sph{S} {Q^\Sph{s}}_\Sph{c} + q\R  D_\Sph{c} A -{Q^\Sph{s}}_\Sph{c}D_\Sph{s}A
			\\
			&+\overset{\star}{k}p_\Sph{c} + AqD_\Sph{c}\R-\frac{1}{2}qD_\Sph{c}A,
		\end{split}
		\label{FinalSystemDiffMom}
	\end{align}
	\newcommand{\ParabolicHyperbolicA}{\Eqsref{ParabolicEquation}--\eqref{FinalSystemDiffMom}\xspace}
	where, $F$ takes the same form as in \Eqref{eq:defF} and $E$ becomes
	\begin{align}
		E&={}^{(2)}R-2p^\Sph{c}p_\Sph{c}-Q_\Sph{cs}Q^\Sph{cs}+\left( 2\R + \frac{1}{2} \right) q^2.
	\end{align} 
	We refer to these equations as the \emph{modified parabolic-hyperbolic system} while \ParabolicHyperbolicR{} is often labelled as the \emph{original parabolic-hyperbolic system}.
	
	While \Eqref{eq:alternativedatadec} looks like a minor modification, it has dramatic consequences for the asymptotics of the solutions \cite{Beyer:2020HW} because of the different way the free data for these equations are specified, see below. First observe that this modification has changed some of the principal part of the system. While the principal part of \Eqref{ParabolicEquation} is unchanged (and is therefore parabolic provided \Eqref{eq:parabolcond} holds as before), the subsystem \Eqsref{FinalSystemDiffNorm} -- \eqref{FinalSystemDiffMom} turns out to be symmetrisable hyperbolic with symmetriser 
	\begin{align}
		\left(
		\begin{array}{cc}
			\frac{1}{2}+ \R & 0 \\
			0 & h^\Sph{ab}
		\end{array}
		\right) 
	\end{align}
	provided 
	\begin{align}
		\frac{1}{2}+\R>0,
		\label{HyperbolicityCondition}
	\end{align}
	where $h^\Sph{ab}$ is the intrinsic inverse of $h_\Sph{ab}$. We  refer to \Eqref{HyperbolicityCondition} as the \textit{hyperbolicity condition}.
	It therefore turns out that \ParabolicHyperbolicA{} is parabolic-hyperbolic provided \Eqsref{eq:parabolcond} and \eqref{HyperbolicityCondition} hold. 
	
	The structure of these equations suggest to group the various fields as follows:
	\begin{description}
		\item[Free data:] The fields $B_\Sph{a}$, $Q_\Sph{ab}$, $h_\Sph{ab}$ and $\R$ are {free data} everywhere on $\Sigma$. 
		\item[Unknowns:] The fields $A$, $q$ and $p_\Sph{a}$ are the unknowns. 
		\item[Cauchy data:] The initial values of the fields $A,q$ and $p_\Sph{c}$ on some $\rho=\rho_0$-surface is the Cauchy data.
	\end{description}
	
	It is important to notice that similar to R\'acz's ``original'' parabolic-hyperbolic system \ParabolicHyperbolicR{}, the PDE conditions \Eqsref{eq:parabolcond} and \eqref{HyperbolicityCondition} are properties of the free data alone and can therefore be verified \emph{before} solutions are constructed.
	In particular, if these conditions for the free data are met, the \emph{Cauchy problem} in the increasing $\rho$-direction is well-posed. We briefly comment on the claim in \cite{Csukas:2020} that it is sufficient to interpret our modified formulation \ParabolicHyperbolicA of the vacuum constraints (introduced in \cite{Beyer:2020HW})  as the special case of R\'acz's ``original'' formulation \ParabolicHyperbolicR where the free field $\kappa$ is chosen to be determined by \Eqref{eq:alternativedatadec} in terms of some given field $\R$ and the unknown $q$ ``on the fly'' at each time step of the evolution; see Section~4.3.1 in \cite{Csukas:2020}. While this claim is evident on the one hand (because both the modified and the original formulations represent the same Einstein vacuum constraints), it may also be misleading. The reason is that this point of view neglects the significant role played by the new hyperbolicity condition \Eqref{HyperbolicityCondition} implied by the  new principal part of the resulting PDEs. Indeed it is possible to construct numerical examples which do not converge when \Eqref{HyperbolicityCondition} is violated.

	\subsubsection{An algebraic-hyperbolic formulation of the constraints}
	\label{Sec:An_algebraic_hyperbolic_formulation_of_the_constraints}
	
	We end this subsection by discussing R\'acz's algebraic-hyperbolic formulation of the constraints \cite{Racz:2015gb}.
	Based on the same $2+1$-framework discussed in \Sectionref{SubSec:Racz2Plus1Decomp}, we now write the Hamiltonian constraint as the following algebraic equation to determine the quantity $\kappa$
	\begin{equation}
		\label{Eq:kappa_AlgHyp}
		\kappa=\frac{1}{2q}\left( -^{(3)}R+Q^\Sph{ab}Q_\Sph{ab}-\frac{1}{2}q^{2}+2p_\Sph{c}p^\Sph{c} \right),	
	\end{equation}
	instead of as the PDE \eqref{ParabolicEquation_Racz} to determine $A$. To this end we consider $A$ as a free field now (as opposed to the unknown  in R\'acz's parabolic-hyperbolic formulation), while $\kappa$ is now determined algebraically from the other fields by \Eqref{Eq:kappa_AlgHyp} (and is therefore not anymore interpreted as the free field of R\'acz's parabolic-hyperbolic formulation).           	
	The equations resulting from this are 
	\begin{gather}
		\begin{split}
			\mathcal L_{\rho} p_\Sph c-B^\Sph a D_\Sph {a}p_\Sph {c} &+\frac{A\kappa}{q}D_\Sph {c}q -\frac{2A}{q}p_\Sph {a}D_\Sph {c}p^\Sph {a} =\,p_\Sph a D_\Sph b B^\Sph a +\overset{\star}{k}p_\Sph c+ \kappa  D_\Sph c A	
			\\
			&- {Q^\Sph a}_\Sph{c}D_\Sph{a}A
			- \frac{1}{2}qD_\Sph{c}A -AD_\Sph{a} {Q^\Sph a}_\Sph{c}+\frac{A}{2q}D_\Sph{c}(-^{(3)}R+Q_\Sph{ab}Q^\Sph{ab}),
			\label{Eqs:AlgHyp_p}
		\end{split}
		\\
		\mathcal{L}_{\rho}q-B^\Sph{a}D_\Sph{a}q-AD_\Sph{a}p^\Sph{a}=\kstar{}^\Sph{ab}Q_\Sph{ab}+\frac{1}{2}q\kstar -\kstar \kappa + 2p_\Sph{c}D^\Sph{c}A,
		\label{Eqs:AlgHyp_q}
	\end{gather}  	
	where $\prescript{(3)}{}{R}$ is given by \Eqref{Eq:R3} and $\kappa$ by \Eqref{Eq:kappa_AlgHyp}.	We refer to this as the \emph{algebraic-hyperbolic formulation} of the Einstein vacuum constraints.
	
	\newcommand{\AlgebraicHyperbolic}{\Eqsref{Eq:kappa_AlgHyp}--\eqref{Eqs:AlgHyp_q}}
	
	According to \cite{Racz:2014kk}, the system \AlgebraicHyperbolic{} is symmetrisable hyperbolic with symmetriser 
	\begin{align}
		\left(
		\begin{array}{cc}
			-\frac{\kappa}{q} & 0 \\
			0 & h^\Sph{ab}
		\end{array}
		\right)
	\end{align} 
	provided 
	\begin{align}
		\kappa q< 0. 
		\label{HyperbolicityCondition_2}
	\end{align} 
	We refer to \Eqref{HyperbolicityCondition_2} as the \textit{algebraic-hyperbolicity condition}. This should not be confused with the hyperbolicity condition \Eqref{HyperbolicityCondition} associated with our modified parabolic-hyperbolic system in \Sectionref{Sec:A_modified_parabolic_hyperbolic_formulation_of_the_constraints}.
	
	All of this suggests to group the $(2+1)$-fields in the following way for this formulation:
	\begin{description}
		\item[Free data:] The fields $B_\Sph{a}$, $Q_\Sph{ab}$, $h_\Sph{a b}$ and $A$ are the free data everywhere on $\Sigma$.
		\item[Unknowns:] The quantities $q$ and $p_ {A}$ are considered as the unknowns.
		\item[Cauchy data:] The initial values of the fields $q$ and $p_\Sph{A}$ on some $\rho=\rho_0$-surface is the Cauchy data.
	\end{description}
	
	It follows that for arbitrary free data and Cauchy data, for which the algebraic-hyperbolicity condition \Eqref{HyperbolicityCondition_2} holds on the initial leaf $\rho=\rho_0$, \AlgebraicHyperbolic{} is a hyperbolic system, and, the {Cauchy problem} (in both the increasing and decreasing $\rho$-directions) is well-posed. Note, however that the algebraic-hyperbolicity condition \Eqref{HyperbolicityCondition_2} is not just a property of the free data. Due to its depends on the unknowns, it is possible to fail  during the evolution even if it holds initially.
	
	\subsection{Remarks about other evolutionary formulations of the constraints}
	\label{sec:otherformulations}
	
	Given the results of \Sectionsref{SubSec:Hyperboloidal_Sphereical} and \ref{Sec:Solving_the_constraints_on_asymptotically_spherical_backgrounds} it is natural to wonder whether or not is possible to construct asymptotically hyperboloidal initial data sets using the other evolutionary formulations of the constraints (discussed in \Sectionref{Sec:Other_evolutionary_formulations_of_the_constraints}). This is exactly the issue that we address in the present subsection. In particular, we provide some evidence that the two evolutionary formulations presented in \Sectionref{Sec:Other_evolutionary_formulations_of_the_constraints} \emph{are not well suited to the construction of asymptotically hyperboloidal initial data sets}.

	\subsubsection{Algebraic-hyperbolic formulation} 
	We first discuss the algebraic-hyperbolic formulation, introduced in \Sectionref{Sec:An_algebraic_hyperbolic_formulation_of_the_constraints}. Suppose  that we had constructed a solution of the algebraic-hyperbolic constraints \AlgebraicHyperbolic{} that is asymptotically hyperboloidal in accordance with \Propref{Result:Hyp_Decays}. As a consequence  we would have
	\begin{align}
		\kappa q = 2\left(\kappa^{(0)}\right)^2 + \DecayO{\rho}{}.
	\end{align}
	It is clear then that the condition \eqref{HyperbolicityCondition_2} $\kappa q <0$ would be violated for all sufficiently large $\rho$. In particular, we conclude that the \emph{algebraic-hyperbolic formulation does not have a well-posed Cauchy problem near $\rho=\infty$ in the asymptotically hyperboloidal setting}. This is consistent with the findings in \cite{mastersthesis}. We shall therefore not discuss this form\-ulation any further.

	\subsubsection{Modified parabolic-hyperbolic formulation} 
	Let us now consider the modified parabolic-hyperbolic formulation presented in \Sectionref{Sec:A_modified_parabolic_hyperbolic_formulation_of_the_constraints}. According to \cite{Beyer:2020HW} the fields 
	\begin{align}
		\R = \frac{(2-V)\rho}{4\left( 1- V \right)}\frac{\partial_\rho V}{V},
		\quad
		\quad
		B_{A}=0,
		\quad
		Q_\Sph{ab}=0,
		\quad 
		h_\Sph{ab}=\rho^{2}\Omega_\Sph{ab}
		\label{Eq:R_Sphere}
	\end{align}
	and 
	\begin{align}
		q=\frac{2\C\, V}{\rho\sqrt{1- V }},
		\quad
		A=\sqrt{\frac{( 1 - V )\rho}{\rho -2m- ( \rho - 2m )V + \rho\, \C^2 V^2 }},
		\quad
		p_\Sph{a}=0,
		\label{FlatGeneralSol_R_A}
	\end{align}   
	constitute a spherically symmetric solution of the modified parabolic-hyperbolic system \ParabolicHyperbolicA, where $\C$ is a free constant and $m$ is the mass.
	
	Suppose now that we set $V=1-\V/\rho^2$, for some constant $\V>0$. Then, the solutions \Eqref{FlatGeneralSol_R_A} have asymptotic radial expansions 
	\begin{align}
		\begin{split}
			A=\frac{\sqrt{\V}}{|\C|\rho}+\DecayO{\rho}{3},\quad q =\frac{\C}{\sqrt{\V}}-\frac{\sqrt{\V}\C}{\rho^{2}},\quad\kappa= \R q =\frac{\C}{2\sqrt{\V}} + \frac{\C \sqrt{\V} }{2\rho^2},
		\end{split}
	\end{align}
	and \Eqref{Eq:R_Sphere} gives
	\begin{align}
		\R = \frac{\rho^{2}+\V}{2\rho^{2}-2 \V}\implies \frac{1}{2}+\R = \frac{\rho^{2}}{\rho^{2}-\V}.
		\label{R_Hyp_Mod}
	\end{align}
	It is a consequence of \Propref{Result:Hyp_Decays} the associated initial data set is therefore asymptotically hyperboloidal. 
	
	An interesting particular solution is now obtained by setting the Cauchy data equal to the values of the background data set at $\rho=\rho_0$. A straightforward calculation shows that the parameter values $\C$ and $m$ corresponding to this particular solution of the vacuum constraints are
	\begin{equation}
		\C =1, \quad m=\frac{1}{2}\left( \frac{\V-\rho_{0}^2}{\rho_0} \right).
	\end{equation}    
	We conclude that if the  vacuum initial data set resulting from this is supposed to have a non-negative Bondi mass, we must pick $\V - \rho_{0}^{2}\ge 0$. However, if this is true then, from \Eqref{R_Hyp_Mod}, we get $1/2+\R<0$ (at $\rho=\rho_0$) and hence we conclude that the hyperbolicity condition \Eqref{HyperbolicityCondition} is violated. Again we conclude that the
	\emph{modified parabolic-hyperbolic formulation does in general not have well-posed Cauchy problem in the asymptotically hyperboloidal setting}. For this reason, we shall not discuss this formulation any further here.
	\end{appendices}


\newpage
	
	\begin{thebibliography}{10}
		
		\bibitem{FouresBruhat:1952ji}
		Yvonne {Four{\`e}s-Bruhat}.
		\newblock Th{\'e}or{\`e}me d'existence pour certains syst{\`e}mes
		d'{\'e}quations aux d{\'e}riv{\'e}es partielles non lin{\'e}aires.
		\newblock {\em Acta Math.}, 88(1):141--225, 1952.
		\newblock DOI:~\href{https://doi.org/10.1007/BF02392131}{10.1007/BF02392131}.
		
		\bibitem{ChoquetBruhat:1969cl}
		Yvonne {Choquet-Bruhat} and Robert~P Geroch.
		\newblock Global aspects of the {{Cauchy}} problem in general relativity.
		\newblock {\em Commun. Math. Phys.}, 14(4):329--335, 1969.
		\newblock DOI:~\href{https://doi.org/10.1007/BF01645389}{10.1007/BF01645389}.
		
		\bibitem{bartnik2004}
		Robert~A Bartnik and James Isenberg.
		\newblock The {{Constraint Equations}}.
		\newblock In {\em The {{Einstein Equations}} and the {{Large Scale Behavior}}
			of {{Gravitational Fields}}}, pages 1--38. {Birkh{\"a}user Physics}, 2004.
		
		\bibitem{Baumgarte:2010vs}
		Thomas~W Baumgarte and Stuart~L Shapiro.
		\newblock {\em Numerical {{Relativity}}}.
		\newblock Solving {{Einstein}}'s {{Equations}} on the {{Computer}}. {Cambridge
			University Press}, 2010.
		
		\bibitem{dilts2017}
		James Dilts, Michael Holst, Tamara Kozareva, and David Maxwell.
		\newblock Numerical {{Bifurcation Analysis}} of the {{Conformal Method}}.
		\newblock 2017.
		\newblock Preprint. \href{http://arxiv.org/abs/1710.03201}{arXiv:1710.03201}.
		
		\bibitem{anderson2018a}
		Michael~T. Anderson.
		\newblock On the conformal method for the {{Einstein}} constraint equations.
		\newblock 2018.
		\newblock Preprint. \href{http://arxiv.org/abs/1812.06320}{arXiv:1812.06320}.
		
		\bibitem{Bishop:1998cb}
		Nigel~T Bishop, Richard Isaacson, Manoj Maharaj, and Jeffrey Winicour.
		\newblock {Black hole data via a Kerr-Schild approach}.
		\newblock {\em Phys. Rev. D}, 57(10):6113--6118, 1998.
		\newblock
		DOI:~\href{https://doi.org/10.1103/PhysRevD.57.6113}{10.1103/PhysRevD.57.6113}.
		
		\bibitem{Matzner:1998hv}
		Richard~A Matzner, Mijan~F Huq, and Deirdre Shoemaker.
		\newblock {Initial data and coordinates for multiple black hole systems}.
		\newblock {\em Phys. Rev. D}, 59(2):024015, 1998.
		\newblock
		DOI:~\href{https://doi.org/10.1103/PhysRevD.59.024015}{10.1103/PhysRevD.59.024015}.
		
		\bibitem{Moreno:2002dm}
		Claudia Moreno, Dar{\'\i}o N{\'u}{\~n}ez, and Olivier Sarbach.
		\newblock {Kerr{\textendash}Schild-type initial data for black holes with
			angular momenta}.
		\newblock {\em Class. Quantum Grav.}, 19(23):6059--6073, 2002.
		\newblock
		DOI:~\href{https://doi.org/10.1088/0264-9381/19/23/312}{10.1088/0264-9381/19/23/312}.
		
		\bibitem{Bishop:2004gb}
		Nigel~T Bishop, Florian Beyer, and Michael Koppitz.
		\newblock {Black hole initial data from a nonconformal decomposition}.
		\newblock {\em Phys. Rev. D}, 69(6):325, 2004.
		\newblock
		DOI:~\href{https://doi.org/10.1103/PhysRevD.69.064010}{10.1103/PhysRevD.69.064010}.
		
		\bibitem{Racz:2014kk}
		Istv{\'a}n R{\'a}cz.
		\newblock Cauchy problem as a two-surface based `geometrodynamics'.
		\newblock {\em Class. Quantum Grav.}, 32(1):015006, 2015.
		\newblock
		DOI:~\href{https://doi.org/10.1088/0264-9381/32/1/015006}{10.1088/0264-9381/32/1/015006}.
		
		\bibitem{Racz:2014dx}
		Istv{\'a}n R{\'a}cz.
		\newblock Is the {{Bianchi}} identity always hyperbolic?
		\newblock {\em Class. Quantum Grav.}, 31(15):155004, 2014.
		\newblock
		DOI:~\href{https://doi.org/10.1088/0264-9381/31/15/155004}{10.1088/0264-9381/31/15/155004}.
		
		\bibitem{Racz:2015gb}
		Istv{\'a}n R{\'a}cz.
		\newblock Constraints as evolutionary systems.
		\newblock {\em Class. Quantum Grav.}, 33(1):015014, 2016.
		\newblock
		DOI:~\href{https://doi.org/10.1088/0264-9381/33/1/015014}{10.1088/0264-9381/33/1/015014}.
		
		\bibitem{Racz:2015bu}
		Istv{\'a}n R{\'a}cz and Jeffrey Winicour.
		\newblock Black hole initial data without elliptic equations.
		\newblock {\em Phys. Rev. D}, 91(12):124013, 2015.
		\newblock
		DOI:~\href{https://doi.org/10.1103/PhysRevD.91.124013}{10.1103/PhysRevD.91.124013}.
		
		\bibitem{Szabados:2009ig}
		L{\'a}szl{\'o}~B Szabados.
		\newblock {Quasi-Local Energy-Momentum and Angular Momentum in General
			Relativity}.
		\newblock {\em Living Rev. Relativity}, 12(4):76, 2009.
		\newblock
		DOI:~\href{https://doi.org/10.1088/0264-9381/14/1a/016}{10.1088/0264-9381/14/1a/016}.
		
		\bibitem{Cerebaum:2016}
		Carla Cederbaum, Julien Cortier, and Anna Sakovich.
		\newblock {On the center of mass of asymptotically hyperbolic initial data
			sets}.
		\newblock {\em Annales Henri Poincaré}, 17(6):1504--1528, 2016.
		\newblock
		DOI:~\href{https://doi.org/10.1007/s00023-015-0438-5}{10.1007/s00023-015-0438-5}.
		
		\bibitem{Beyer:2017tu}
		Florian Beyer, Leon Escobar, and J{\"o}rg Frauendiener.
		\newblock {Asymptotics of solutions of a hyperbolic formulation of the
			constraint equations}.
		\newblock {\em Class. Quantum Grav.}, 34(20):205014, 2017.
		\newblock
		DOI:~\href{https://doi.org/10.1088/1361-6382/aa8be6}{10.1088/1361-6382/aa8be6}.
		
		\bibitem{Beyer:2018HW}
		Florian Beyer, Leon Escobar, J{\"o}rg Frauendiener, and Joshua Ritchie.
		\newblock Numerical construction of initial data sets of binary black hole type
		using a parabolic-hyperbolic formulation of the vacuum constraint equations.
		\newblock {\em Class. Quantum Grav.}, 36(17):175005, 2019.
		\newblock
		DOI:~\href{https://doi.org/10.1088/1361-6382/ab3482}{10.1088/1361-6382/ab3482}.
		
		\bibitem{Beyer:2020HW}
		J{\"o}rg~Frauendiener Florian~Beyer and Joshua Ritchie.
		\newblock Asymptotically flat vacuum initial data sets from a modified
		parabolic-hyperbolic formulation of the {E}instein vacuum constraint
		equations.
		\newblock {\em Phys. Rev. D}, 101:084013, Apr 2020.
		\newblock
		DOI:~\href{https://link.aps.org/doi/10.1103/PhysRevD.101.084013}{10.1103/PhysRevD.101.084013}.
		
		\bibitem{Csukas:2020}
		K{\'a}roly Csuk{\'a}s and Istv{\'a}n R{\'a}cz.
		\newblock Numerical investigations of the asymptotics of solutions to the
		evolutionary form of the constraints.
		\newblock {\em Class. Quantum Grav.}, 37(15):155006, 2020.
		\newblock
		DOI:~\href{https://doi.org/10.1088/1361-6382/ab8fce}{10.1088/1361-6382/ab8fce}.
		
		\bibitem{Beyer:2017jw}
		Florian Beyer, Leon Escobar, and J{\"o}rg Frauendiener.
		\newblock Criticality of inhomogeneous {{Nariai}}-like cosmological models.
		\newblock {\em Phys. Rev. D}, 95(8):084030, 2017.
		\newblock
		DOI:~\href{https://doi.org/10.1103/PhysRevD.95.084030}{10.1103/PhysRevD.95.084030}.
		
		\bibitem{Beyer:2014bu}
		Florian Beyer, Boris Daszuta, J{\"o}rg Frauendiener, and Ben Whale.
		\newblock Numerical evolutions of fields on the 2-sphere using a spectral
		method based on spin-weighted spherical harmonics.
		\newblock {\em Class. Quantum Grav.}, 31(7):075019, 2014.
		\newblock
		DOI:~\href{https://doi.org/10.1088/0264-9381/31/7/075019}{10.1088/0264-9381/31/7/075019}.
		
		\bibitem{Beyer:2016fc}
		Florian Beyer, Leon Escobar, and J{\"o}rg Frauendiener.
		\newblock Numerical solutions of {{Einstein}}'s equations for cosmological
		spacetimes with spatial topology {$S^3$} and symmetry group {$U(1)$}.
		\newblock {\em Phys. Rev. D}, 93(4):043009, 2016.
		\newblock
		DOI:~\href{https://doi.org/10.1103/PhysRevD.93.043009}{10.1103/PhysRevD.93.043009}.
		
		\bibitem{Beyer:2009vw}
		Florian Beyer.
		\newblock A spectral solver for evolution problems with spatial
		{$S^3$}-topology.
		\newblock {\em J. Comp. Phys.}, 228(17):6496--6513, 2009.
		\newblock
		DOI:~\href{https://doi.org/10.1016/j.jcp.2009.05.037}{10.1016/j.jcp.2009.05.037}.
		
		\bibitem{Alcubierre:Book}
		Miguel Alcubierre.
		\newblock {\em Introduction to 3+1 {{Numerical Relativity}}}.
		\newblock {Oxford Science Publications}, 2008.
		
		\bibitem{Penrose:1984tf}
		Roger Penrose and Wolfgang Rindler.
		\newblock {\em Two-{{Spinor Calculus}} and {{Relativistic Fields}}}, volume~1
		of {\em Spinors and {{Space}}-{{Time}}}.
		\newblock {Cambridge University Press}, {Cambridge}, 1984.
		
		\bibitem{Beyer:2015bv}
		Florian Beyer, Boris Daszuta, and J{\"o}rg Frauendiener.
		\newblock A spectral method for half-integer spin fields based on spin-weighted
		spherical harmonics.
		\newblock {\em Class. Quantum Grav.}, 32(17):175013, 2015.
		\newblock
		DOI:~\href{https://doi.org/10.1088/0264-9381/32/17/175013}{10.1088/0264-9381/32/17/175013}.
		
		\bibitem{PhD:HypDef}
		Lars Andersson and Piotr~T. Chru{\'s}ciel.
		\newblock Solutions of the constraint equations in general relativity
		satisfying ``hyperboloidal boundary conditions".
		\newblock {\em Department of Mathematics, Royal Institute of Technology, S10044
			Stockholm, Sweden}, 1996.
		\newblock
		URL:~\href{https://homepage.univie.ac.at/piotr.chrusciel/papers/Dissertationes/Dissertationes.pdf}{homepage.univie.ac.at}.
		
		\bibitem{penrose1986}
		Roger Penrose and Wolfgang Rindler.
		\newblock {\em Spinor and {{Twistor Methods}} in {{Space}}-{{Time Geometry}}},
		volume~2 of {\em Spinors and {{Space}}-{{Time}}}.
		\newblock {Cambridge University Press}, 1986.
		
		\bibitem{mastersthesis}
		{Joshua Ritchie}.
		\newblock Asymptotics of solutions in evolutionary formulations of the
		{E}instein constraint equations.
		\newblock Master's thesis, University of Otago, Febuary 2018.				
		
	\end{thebibliography}
\end{document}